\input{config/fixes}

\documentclass[a4paper,11pt]{article}

\usepackage{jcappub}

\usepackage[UKenglish]{babel}
\usepackage{bbm}
\usepackage[numbered]{bookmark}
\usepackage{booktabs}
\usepackage[inline]{enumitem}
\usepackage{etoolbox}
\usepackage[T1]{fontenc}
\usepackage{journalref}
\usepackage{lmodern}
\usepackage{mathrsfs}
\usepackage{mathtools}
\usepackage{microtype}
\usepackage{multirow}
\usepackage{physics}
\usepackage{placeins}
\usepackage{ragged2e}
\usepackage{siunitx}
\usepackage{textgreek}
\usepackage{upgreek}
\usepackage{xparse}
\usepackage{cleveref}

\hypersetup{
	pdfauthor={M. S. Wang, S. Avila, D. Bianchi, R. Crittenden, W. J. Percival},
	pdfsubject={large-scale structure cosmology},
    pdfkeywords={cosmological parameters from LSS, galaxy clustering, power spectrum, redshift surveys}
}

\graphicspath{{graphics/}}

\DeclareFontFamily{U}{mathx}{\hyphenchar\font45}
\DeclareFontShape{U}{mathx}{m}{n}{
    <5> <6> <7> <8> <9> <10>
    <10.95> <12> <14.4> <17.28> <20.74> <24.88>
    mathx10
}{}
\DeclareSymbolFont{mathx}{U}{mathx}{m}{n}

\DeclareMathAccent{\widebar}{0}{mathx}{"73}
\DeclareMathAccent{\widehat}{0}{mathx}{"70}
\DeclareMathAccent{\widetilde}{0}{mathx}{"72}

\setlist[enumerate]{label=\arabic*)}

\DeclareSIUnit\parsec{pc}
\DeclareSIUnit\h{\text{$h$}}
\DeclareSIUnit\solarmass{\text{$M_\odot$}}

\appto\bibfont{\RaggedRight}

\preto\subequations{\ifhmode\unskip\fi}

\DeclareDocumentCommand{\ComplexNormal}{s m}{%
    \operatorname{\mathbb{C}N}%
    \IfBooleanTF{#1}{(#2)}{\mathopen{}\left(#2\right)\mathclose{}}%
}
\DeclareDocumentCommand{\Cov}{s o m}{%
    \operatorname{cov}%
    \IfBooleanTF{#1}{#2}{%
        \IfNoValueTF{#2}{\mathopen{}\left(#3\right)\mathclose{}}{\mathopen{}#2(#3#2)\mathclose{}}%
    }%
}
\DeclareDocumentCommand{\Diag}{s o m}{%
    \operatorname{diag}%
    \IfBooleanTF{#1}{#3}{%
        \IfNoValueTF{#2}{\mathopen{}\left(#3\right)\mathclose{}}{\mathopen{}#2(#3#2)\mathclose{}}%
    }%
}
\DeclareDocumentCommand{\Exp}{s o m}{%
    \operatorname{\mathbb{E}}%
    \IfBooleanTF{#1}{#3}{%
        \IfNoValueTF{#2}{\mathopen{}\left(#3\right)\mathclose{}}{\mathopen{}#2(#3#2)\mathclose{}}%
    }%
}
\DeclareDocumentCommand{\Like}{s o o m}{%
    \IfNoValueTF{#2}{\operatorname{\mathscr{L}}}{\operatorname{\mathscr{L}_{#2}}}%
    \IfBooleanTF{#1}{#4}{%
        \IfNoValueTF{#3}{\mathopen{}\left(#4\right)\mathclose{}}{\mathopen{}#3(#4#3)\mathclose{}}%
    }%
}
\DeclareDocumentCommand{\Normal}{s o m}{%
    \operatorname{N}%
    \IfBooleanTF{#1}{#3}{%
        \IfNoValueTF{#2}{\mathopen{}\left(#3\right)\mathclose{}}{\mathopen{}#2(#3#2)\mathclose{}}%
    }%
}
\DeclareDocumentCommand{\Prob}{s o m}{%
    \operatorname{\mathbb{P}}%
    \IfBooleanTF{#1}{#2}{%
        \IfNoValueTF{#2}{\mathopen{}\left(#3\right)\mathclose{}}{\mathopen{}#2(#3#2)\mathclose{}}%
    }%
}
\DeclareDocumentCommand{\Var}{s o m}{%
    \operatorname{var}%
    \IfBooleanTF{#1}{#2}{%
        \IfNoValueTF{#2}{\mathopen{}\left(#3\right)\mathclose{}}{\mathopen{}#2(#3#2)\mathclose{}}%
    }%
}
\DeclareDocumentCommand{\given}{s o m m}{%
    \IfBooleanTF{#1}{#3\kern0.5pt\vert\kern0.5pt#4}{
        \IfNoValueTF{#2}{\left.#3\kern0.5pt\middle\vert\kern0.5pt#4\right.}{#3\kern0.5pt#2\vert\kern0.5pt#4}%
    }%
}
\DeclareDocumentCommand{\set}{s O{} m}{%
    \IfBooleanTF{#1}{\lbrace#3\rbrace}{
        \IfNoValueTF{#2}{\mathopen{}\qty{#3}\mathclose{}}{\mathopen{}#2\lbrace#3#2\rbrace\mathclose{}}%
    }
}

\newcommand{\orcid}[1]{%
    \hspace{0.3\baselineskip}%
    \raisebox{-1pt}{\href{https://orcid.org/#1}{
        \includegraphics[height=0.75\baselineskip,keepaspectratio]{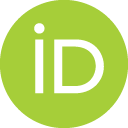}
    }}%
    \hspace{-0.5ex}%
}
\newcommand{\preprint}[1]{\normalfont{preprint}}
\newcommand{\is}{is}
\newcommand{\codename}[1]{\textsc{#1}}

\makeatletter
    \let\olduppi\uppi
    \def\uppi{\mkern 1mu\olduppi\mkern 1mu}
\makeatother

\renewcommand{\max}{\mathrm{max}}

\newcommand{\e}{\mathrm{e}}
\newcommand{\im}{\mkern 1mu\mathrm{i}\mkern 1mu}
\newcommand{\R}{\mathbb{R}}
\newcommand{\C}{\mathbb{C}}

\newcommand{\conj}[1]{#1^*}
\newcommand{\mat}[1]{\mathsf{#1}}
\newcommand{\trans}[1]{#1^{\intercal}}
\newcommand{\herm}[1]{#1^{\dagger}}
\newcommand{\est}[1]{\widehat{#1}}
\newcommand{\filter}[1]{\widetilde{#1}}
\newcommand{\placedot}{\operatorname{\vdot}}
\newcommand{\measurement}[3]{{#1}^{#2}_{#3}}

\newcommand{\dirac}{\delta^{\textrm{D}}}
\newcommand{\kron}{\delta^{\textrm{K}}}
\newcommand{\legendre}{\mathcal{L}}
\newcommand{\sphY}[1]{Y_{#1}}
\newcommand{\sphYc}[1]{\conj{Y}_{#1}}

\newcommand{\fNL}{f_{\textrm{NL}}}
\newcommand{\fsky}{f_{\textrm{sky}}}
\newcommand{\nbar}{\bar{n}}
\newcommand{\den}{\delta}

\newcommand{\los}{\vu{n}}
\newcommand{\vr}{\vb{r}}
\newcommand{\vs}{\vb{s}}
\newcommand{\vv}{\vb{v}}
\newcommand{\vDelta}{\vb{\Delta}}
\newcommand{\vk}{\vb{k}}
\newcommand{\vq}{\vb{q}}
\newcommand{\vur}{\vu{r}}
\newcommand{\vus}{\vu{s}}

\newcommand{\vuk}{\vu{k}}

\newcommand{\matter}{{\textrm{m}}}
\newcommand{\shot}{\textrm{shot}}
\newcommand{\gal}{{\textrm{g}}}
\newcommand{\syn}{{\textrm{s}}}
\newcommand{\true}{\textrm{true}}
\newcommand{\fid}{\textrm{fid}}
\newcommand{\angular}{\textrm{A}}
\newcommand{\sound}{\textrm{s}}
\newcommand{\drag}{\textrm{d}}
\newcommand{\sph}{\textrm{sph}}
\newcommand{\Cart}{\textrm{Cart}}
\newcommand{\hyb}{\textrm{hyb}}
\newcommand{\data}{\textrm{d}}

\title{Hybrid-basis inference for large-scale galaxy clustering: combining spherical and Cartesian Fourier analyses}

\author[a]{M.~S.~Wang\orcid{0000-0002-2652-4043},}
\author[b,c]{S.~Avila\orcid{0000-0001-5043-3662},}
\author[d]{D.~Bianchi\orcid{0000-0001-9712-0006},}
\author[a]{R.~Crittenden\orcid{0000-0002-5743-1528}}
\author[e,f,g]{and W.~J.~Percival\orcid{0000-0002-0644-5727}}

\affiliation[a]{Institute of Cosmology and Gravitation, University of Portsmouth \\ Burnaby Road, Portsmouth, Hampshire PO1 3FX, United Kingdom}
\affiliation[b]{Instituto de F\'{i}sica Te\'{o}rica UAM-CSIC \\ Nicol\'{a}s Cabrera, 13--15, 28049 Madrid, Spain}
\affiliation[c]{Departamento de F\'{i}sica Te\'{o}rica, Facultad de Ciencias, Universidad Aut\'{o}noma de Madrid \\ Cantoblanco, 28049 Madrid, Spain}
\affiliation[d]{ICC, University of Barcelona, IEEC-UB \\ Mart\'{i} i Franqu\'{e}s, 1, E-08028 Barcelona, Spain}
\affiliation[e]{Waterloo Centre for Astrophysics, University of Waterloo \\ 200 University Avenue West, Waterloo, Ontario N2L 3G1, Canada}
\affiliation[f]{Department of Physics and Astronomy, University of Waterloo \\ 200 University Avenue West, Waterloo, Ontario N2L 3G1, Canada}
\affiliation[g]{Perimeter Institute for Theoretical Physics \\ 31 Caroline Street North, Waterloo, Ontario N2L 2Y5, Canada}

\emailAdd{mike.wang@port.ac.uk}
\emailAdd{santiago.avila@uam.es}
\emailAdd{davide.bianchi@icc.ub.edu}
\emailAdd{robert.crittenden@port.ac.uk}
\emailAdd{will.percival@uwaterloo.ca}

\abstract{Future precision cosmology from large-scale structure experiments including the Dark Energy Spectroscopic Instrument~(DESI) and \textit{Euclid} will probe wider and deeper cosmic volumes than those covered by previous surveys. The Cartesian power spectrum analysis of anisotropic galaxy clustering based on the Fourier plane wave basis makes a number of assumptions, including the local plane-parallel approximation, that will no longer be valid on very large scales and may degrade cosmological constraints. We propose an approach that util{\is}es a hybrid basis: on the largest scales, clustering statistics are decomposed into spherical Fourier modes which respect the natural geometry of both survey observations and physical effects along the line of sight, such as redshift-space distortions, the Alcock--Paczy\'{n}sky and light-cone effects; on smaller scales with far more clustering modes, we retain the computational benefit of the power spectrum analysis aided by fast Fourier transforms. This approach is particularly suited to the likelihood analysis of local primordial non-Gaussianity $\fNL$ through the scale-dependent halo bias, and we demonstrate its applicability with $N$-body simulations. We also release our public code \codename{harmonia} for galaxy clustering likelihood inference in spherical Fourier or hybrid-basis analyses.}

\keywords{cosmological parameters from LSS, galaxy clustering, power spectrum, redshift surveys}

\arxivnumber{2007.14962}

\begin{document}

\maketitle

\flushbottom

\section{Introduction}
\label{sec:introduction}

In the past decades, the strongest cosmological model constraints have mostly come from full-sky temperature and polar{\is}ation measurements of the cosmic microwave background~(CMB), with the now decommissioned \textit{Planck}\footnote{\href{http://www.esa.int/planck}{\texttt{esa.int/planck}}}~satellite providing percent-level precision on the standard {\textLambda}CDM model parameters~\cite{Planck_2018}. However, as primary CMB probes gradually saturate the cosmic variance bound, large-scale structure~(LSS) observations have become an indispensable and complementary tool for studying the Universe with 3-dimensional data sets. In particular, the clustering of galaxies as tracers of the underlying matter distribution offers powerful methods for measuring the late-time expansion history through baryon acoustic oscillations~(BAOs) as well as the growth of cosmic structure from redshift-space distortions~(RSDs). Recent experiments such as the Baryon Oscillation Spectroscopic Survey\footnote{\href{https://www.sdss.org/surveys/boss/}{\texttt{sdss.org/surveys/boss/}}}~(BOSS) and its extension eBOSS\footnote{\href{https://www.sdss.org/surveys/eboss/}{\texttt{sdss.org/surveys/eboss/}}} have already provided some competitive results to CMB analyses \cite{BOSS_2017,eBOSS_2020}. Upcoming galaxy redshift surveys, such as the Dark Energy Spectroscopic Instrument\footnote{\href{https://desi.lbl.gov}{\texttt{desi.lbl.gov}}}~(DESI) and \textit{Euclid}\footnote{\href{https://www.euclid-ec.org/}{\texttt{euclid-ec.org}}}~\cite{DESI_2016,Euclid_2011}, will map out unprecedented cosmological volumes and thus be able to probe local primordial non-Gaussianity~(PNG) and relativistic effects that leave scale-dependent signatures on cosmological scales approaching the Hubble horizon size; these studies are important for differentiating amongst a plethora of inflationary models and testing the nature of gravity on ultra-large scales~\cite{Dalal_2008,Matarrese_2008,Slosar_2008,Yoo_2009,Bonvin_2011,Challinor_2011,Bruni_2012,Jeong_2012,Camera_2015,Alonso_2015,Fonseca_2015}.

In linear perturbation theory, cosmic fluctuations are well described by Gaussian random fields if the initial conditions are also Gaussian. All statistical information is encoded in the 2-point correlator, which can be described either in configuration space by the correlation function~$\xi(\vr)$ or in Fourier space by the power spectrum~$P(\vk)$, though the latter is better at disentangling fluctuations on different scales. Under the cosmological principle, the underlying matter power spectrum is expected to be spatially homogeneous and isotropic on large scales, but the observed power spectrum of galaxies is not, since their spatial positions have to be inferred from the observed redshift and angular positions in the sky. This process can be affected by both RSDs and the Alcock--Paczy\'{n}ski~(AP) effect, which result in apparent distortions in the galaxy distribution with anisotropies induced around the line of sight~\cite{Kaiser_1987,Alcock_1979}. However, conversely, by measuring galaxy clustering anisotropies, one could in turn constrain the growth rate of structure and cosmological distances.

Following the seminal work by Kaiser~\cite{Kaiser_1987}, it has now become the standard galaxy clustering analysis to adopt the anisotropic galaxy power spectrum~$P(k, \mu)$ or equivalently its Legendre multipoles~$P_\ell(k)$ as the summary statistics, both of which are defined with respect to a fixed global line of sight~$\los$ in the so-called distant-observer and plane-parallel approximations. Furthermore, in more complex non-linear model extensions to redshift-space galaxy clustering~\cite[e.g.][]{Taruya_2010}, $P(k, \mu)$ remains the key observable predicted by perturbation theory. Although these approximations have been justified for past surveys with small sky coverage or for analysis of $N$-body simulations confined to a Cartesian box, the intrinsic spherical geometry of LSS observations with the observer centred at the origin means that $P(k, \mu)$ cannot actually be measured directly. The discrepancy between global plane-parallel predictions of the multipoles and local plane-parallel estimators, which depend on the choice of $\los$ that does vary across the survey volume but force the same $\los$ for pairs of galaxies is known as the wide-angle effect, and it can be a significant systematic on large scales and degrade constraints on cosmological parameters~\cite{Raccanelli_2010,Yoo_2014,Beutler_2019}. It has recently been shown that, even for BOSS-like data sets, wide-angle effects coupled to the survey window can contribute up to \SI{5}{\percent} uncertainties in the even multipoles and dominate the odd multipoles which are important to searches for relativistic effects~\cite{Beutler_2019}.

To deal with this critical issue, new estimators for the 2-point function in configuration space have been proposed~\cite{Szalay_1998,Szapudi_2004,Papai_2008}, and perturbative wide-angle corrections in Fourier space have also been recently derived~\cite{Reimberg_2016,Castorina_2018,Beutler_2019}. Yet despite these efforts, the Cartesian power spectrum analysis still encounters challenges on other fronts. For instance, the analysis is tomographic, which requires fine-tuned binning in redshift; the covariance matrix is essentially a 4-point function and typically intractable analytically, so a large number of mock catalogues are needed for estimation, but this is computationally costly and often does not take cosmology dependence into account~\cite{Eifler_2009,Wang_2019}, and has to allow for errors in the covariance matrix~\cite{Dodelson_2013,Taylor_2013,Percival_2014,Sellentin_2015}. At a more fundamental level, the power spectrum analysis is based on the plane wave basis which forces the survey geometry to align with a Cartesian coordinate system. Given much of the physics affecting galaxy clustering, e.g.~relativistic and light-cone effects, is along the line-of-sight direction, a natural question arises as to whether there is an alternative approach to analysing anisotropic galaxy clustering on large scales that respects the symmetries of the problem~\cite[e.g.][]{Castorina_2018}.

Indeed, such an approach has been proposed before: the 3-dimensional spherical Fourier analysis, also known as spherical harmonic analysis, was first laid down in refs.~\cite{Fisher_1995,Heavens_1995} and then subsequently applied to the Infrared Astronomical Satellite (IRAS) Point Source Catalog Redshift~(PSCz) Survey\footnote{\href{http://irsa.ipac.caltech.edu/Missions/iras.html}{\texttt{irsa.ipac.caltech.edu/Missions/iras.html}}}~\cite{Tadros_1999,Schmoldt_1999,Taylor_2001}, the 2-Micron All-Sky Redshift Survey~(2MASS)\footnote{\href{https://old.ipac.caltech.edu/2mass/}{\texttt{old.ipac.caltech.edu/2mass}}}~\cite{Erdogdu_2006a,Erdogdu_2006b}, and the final catalogue of the 2-Degree Field Galaxy Redshift Survey\footnote{\href{http://www.2dfgrs.net/}{\texttt{2dfgrs.net}}}~(2dFGRS)~\cite{Percival_2004}. The analysis procedure is in general more complex and computationally expensive, and thus less economical for past surveys covering a small sky fraction. In recent years there has been renewed interest in this approach~\cite{Leistedt_2012,Rassat_2012,Shapiro_2012,Pratten_2013,Yoo_2013,Nicola_2014,Lanusse_2015}, owing to the need for a methodology suited for future wide surveys that can probe much large cosmological scales for studying PNG and relativistic effects. However, because of the computational costs and the difficulty in formulating non-linear galaxy clustering models in the spherical Fourier analysis, the Cartesian power spectrum analysis aided by fast Fourier transforms~(FFTs) is currently preferred.

Inspired by the use of hybrid estimators in CMB studies~\cite{Tegmark_1997a,Bond_1998,Bond_2000,Wandelt_2001,Efstathiou_2004,Hinshaw_2007,Hamimeche_2008}, we set out to investigate in this work whether a hybrid-basis approach to likelihood inference from galaxy clustering measurements is viable; this is akin to \textit{Planck}'s low-$\ell$ likelihood directly built from map pixels and high-$\ell$ likelihood based on the compressed pseudo-$C_\ell$ estimator~\cite{Planck_2014,Planck_2016,Planck_2019}. We propose a Fourier analysis that uses spherically decomposed statistics to describe anisotropic clustering on the largest scales in the survey, and switches to the Cartesian power spectrum analysis on comparatively smaller scales. The key advantage of this approach is that while the computational edge of FFTs is util{\is}ed when the clustering modes are numerous, the spherically decomposed clustering statistics can also faithfully capture physical and observational effects parallel and transverse to the line of sight without making any geometric approximations, which impact measurements on very large scales in particular. We present our work as follows:
    \begin{enumerate}
        \item In section~\ref{sec:anisotropic clustering}, we first review both RSD and the AP effects that give rise to anisotropies in galaxy clustering; these are typically described by the power spectrum~$P(k, \mu)$ and equivalently its Legendre multipoles~$P_\ell(k)$ in the distant-observer and global plane-parallel approximations. We also discuss window effects in incomplete survey~observations;
        \item In section~\ref{sec:Cartesian analysis}, we introduce the FFT-based Yamamoto estimator for measuring the power spectrum multipoles~$P_\ell(k)$ in the local plane-parallel approximation \cite{Yamamoto_2006,Bianchi_2015,Scoccimarro_2015,Hand_2017}. We revisit the wide-angle effect and discuss the root cause of the discrepancy between clustering measurements and the theoretical observable~$P(k, \mu)$, which is that the Cartesian plane wave basis does match the inherent geometry of the survey;
        \item Next, we introduce the spherical Fourier analysis in section~\ref{sec:spherical analysis} and extend the original works of refs.~\cite{Fisher_1995,Heavens_1995}. By making comparisons as well as connections between the spherical Fourier and Cartesian power spectrum analyses, we motivate the hybrid-basis approach to analysing galaxy clustering measurements;
        \item In section~\ref{sec:hybrid basis approach}, we lay out the steps involved in constructing the likelihood function in the hybrid-basis analysis, and discuss the technical aspects of likelihood evaluation. We then demonstrate the applicability of our methodology in section~\ref{sec:application} by inferring the local PNG parameter~$\fNL$ from $N$-body simulations with (non-)Gaussian initial conditions. We conclude and motivate future work in section~\ref{sec:discussion}.
    \end{enumerate}

Since the spherical Fourier analysis of galaxy clustering is less common in practical applications than the Cartesian power spectrum analysis, we have made the code used in this work publicly available as a Python package named \textsc{harmonia},\footnote{\href{https://github.com/MikeSWang/Harmonia}{\texttt{github.com/MikeSWang/Harmonia}}\label{footnote:Harmonia}} which could be useful for future work including the survey analysis of upcoming DESI and \textit{Euclid} missions.

\section{Anisotropic galaxy clustering}
\label{sec:anisotropic clustering}

Let us first consider the underlying connection between the galaxy over-density field~$\den(\vr, z)$ and the matter density contrast~$\den_{\matter}(\vr, z)$ as they co-evolve with cosmological redshift~$z$ in real space. At comoving coordinates~$\vr$, the galaxy over-density field is defined by
    \begin{equation}
        \den(\vr, z) = \frac{n(\vr, z) - \nbar(\vr, z)}{\nbar(\vr, z)} \,,
        \label{eq:galaxy over-density field}
    \end{equation}
where $\nbar(\vr, z) = \ev{n(\vr, z)}$ is the ensemble expectation of the observed galaxy number density field~$n(\vr, z)$. In the absence of survey window effects, $\nbar(\vr, z) \equiv \nbar(z)$ is the spatially homogeneous background number density. In the linear perturbative regime, if one ignores lensing or relativistic contributions, then
    \begin{equation}
        \den(\vr, z) = b_k(z) D(z) \den_{\matter,0}(\vr) \,,
        \label{eq:galaxy--matter co-evolution}
    \end{equation}
where $b_k(z)$ is the possibly scale-dependent linear galaxy bias, $D(z)$ is the linear growth factor normal{\is}ed to $D_0 = 1$, and the subscript $\placedot_0$ denotes a quantity evaluated at the present epoch $z = 0$~\cite{Lanusse_2015}. 
Note that since one could only detect galaxies along the \emph{past light-cone}, the comoving coordinates~$\vr$ and redshift~$z$ are \emph{no longer independent} in observations but rather are related by the distance--redshift relationship~$r = \chi(z)$ in a particular cosmological model.

Just like the underlying matter distribution, the clustering of galaxies in real space is expected to be statistically homogeneous and isotropic on large scales under the cosmological principle. However, two sources of anisotropy can be introduced in observations, which we discuss in the following subsections: \emph{redshift-space distortions}, which occur as the inferred line-of-sight position of a galaxy from its redshift is affected by peculiar motions~\cite{Kaiser_1987}; and the \emph{Alcock--Paczy\'{n}sky effect}, which arises when the radial and transverse comoving distances calculated from galaxy redshifts and angular positions in a fiducial cosmological model are rescaled differently compared to the true cosmology~\cite{Alcock_1979,Ballinger_1996}. Moreover,  incomplete survey observations will also introduce spatial variations in the background number density and thus produce window effects in measured clustering statistics, which we discuss in the final~subsection.

Throughout this work, the following convention for the Fourier transform of a field~$f(\vr) \in \R^3$ and its inverse is adopted:
    \begin{equation*}
        f(\vk) = \int \dd[3]{\vr} \e^{-\im\vk\vdot\vr} f(\vr) \,, \quad f(\vr) = \int \frac{\dd[3]{\vk}}{(2\uppi)^3} \e^{\im\vk\vdot\vr} f(\vk) \,.
        \label{eq:Fourier transform convention}
    \end{equation*}
Under periodic boundary conditions imposed on a rectangular box of side lengths~$L_x, L_y, L_z$, the plane wave basis~$\set{\ket{\vk} = \ket{k_x, k_y, k_z}}$ (in bra--ket notation) is discret{\is}ed, with wave numbers being multiples of the fundamental wave number~$\Delta k_{x,y,z} = 2\uppi/L_{x,y,z}$. The inverse transform then becomes an expansion,
    \begin{equation}
        f(\vr) = \frac{1}{V} \sum_{\vk} f(\vk) \braket{\vr}{\vk} \,,
        \label{eq:discrete Fourier expansion}
    \end{equation}
where $V = L_x L_y L_z$~and~$\braket{\vr}{\vk} = \e^{\im\vk\vdot\vr}$. The Dirac and Kronecker delta functions are respectively represented by
    \begin{equation}
        \dirac(\vr - \vr') = \frac{1}{V} \sum_{\vk} \braket{\vr}{\vk} \! \braket*{\vk}{\vr'} \,, \quad \kron_{\vk\vk'} = \frac{1}{V} \int_V \dd[3]{\vr} \braket{\vk}{\vr} \! \braket*{\vr}{\vk'} \,.
        \label{eq:discrete Fourier delta function representations}
    \end{equation}

\subsection{Redshift-space distortions}

For a galaxy that is moving relative to the background expansion of the Universe under the influence of the local gravitational field, its observed redshift~$z^\textrm{obs}$ differs from the cosmological redshift~$z$ owing to the Doppler effect,
    \begin{equation}
        z^\textrm{obs} \simeq z + \frac{\vv \vdot \vur}{a c} \,,
        \label{eq:peculiar velocity Doppler redshift}
    \end{equation}
where $\vv$ is peculiar velocity, $a(z)$ is the scale factor and $c$ is the speed of light~\cite{Peacock_1998}. In the observer's frame of reference, the \emph{redshift-space} coordinates~$\vs$ are mapped from the real-space coordinates $\vr$ along the radial direction by
    \begin{equation}
        s = \chi(z^\textrm{obs}) = r + u
        \label{eq:redshift-space coordinate mapping}
    \end{equation}
to linear order, where $u = \flatfrac{\vv \vdot \vur}{(a H)}$~and~$H(z)$ is the Hubble parameter.

By considering the local conservation of the observed galaxy number density, $n(\vr) \dd[3]{\vr} = n(\vs) \dd[3]{\vs}$, Kaiser first showed that in linear perturbation theory the redshift-space galaxy over-density field is given by~\cite{Kaiser_1987}
    \begin{equation}
        \den(\vs, z) = \qty[b_k(z) + f(z) \mathcal{D}(r)] \den_\matter(\vr, z) \,,
        \label{eq:redshift-space over-density field}
    \end{equation}
where $f(z) = \dv*{\ln D}{\ln a}$ is the linear growth rate and
    \begin{equation}
        \mathcal{D}(r) = \partial_r^2 \nabla^{-2} + \qty[2 + \pdv{\ln\nbar(\vr, z)}{\ln{r}}] \frac{\partial_r}{r} \nabla^{-2}
        \label{eq:RSD operator}
    \end{equation}
is the RSD operator. If we were to Fourier transform the field~$\den(\vs, z)$, the second term in the RSD operator above would introduce mode coupling which complicates the modelling of the redshift-space power spectrum; however, in the \emph{distant-observer approximation} where this mode-coupling term can be neglected, the RSD operator reduces to $\mathcal{D}(r) = \partial_r^2 \nabla^{-2}$~\cite{Kaiser_1987,Zaroubi_1996}. Therefore the galaxy clustering mode in Fourier space is given by\footnote{In this work, we only consider the Fourier modes of galaxy clustering in redshift space, so we will not label $\den(\vk)$ with~$s$ to distinguish it from a real-space quantity. Meanwhile, the Fourier-space matter density contrast~$\den_\matter(\vk)$ always corresponds to the real configuration space.}
    \begin{equation}
        \delta(\vk) = b_k \delta_\matter(\vk) + f \int \dd[3]{\vr} \e^{\im\vk\vdot\vr} \qty\big(\vuk \vdot \vur)^2 \den_\matter(\vr) \,.
        \label{eq:distant-observer approximation}
    \end{equation}
To further simplify this result, the \emph{global plane-parallel approximation} is ubiquitously adopted --- it assumes that the cosine of the angle between the mode vector and the line of sight, $\mu = \vuk \vdot \vur$, becomes independent of $\vur$ if there is a global line of sight~$\los$ for which $\mu \approx \vuk \cdot \los$. The equation above is then reduced to
    \begin{equation}
        \delta(\vk) = \qty(b_k  + f \mu^2) \delta_\matter(\vk) \,,
        \label{eq:global plane-parallel approximation}
    \end{equation}
where the addition of $f \mu^2$ signifies that the effect of RSDs is to introduce a quadrupole anisotropy about the line of sight in the galaxy clustering mode. The redshift-space galaxy power spectrum is therefore 2-dimensional in effect,
    \begin{equation}
        P(\vk) = P(k, \mu) = \qty(b_k  + f \mu^2)^2 P_\matter(k) \,,
        \label{eq:standard Kaiser model}
    \end{equation}
in contrast to the underlying matter power spectrum which is isotropic and thus effectively 1-dimensional.

\subsection{The Alcock--Paczy\'{n}sky effect}

Just as the radial position of galaxies is inferred from their redshift with the comoving distance--redshift relationship, which depends on the Hubble parameter~$H(z)$, the transverse separation between galaxies also needs to be inferred from their angular positions in the sky with the angular diameter distance~$D_\angular(z)$. If one uses a fiducial cosmological model that deviates from the true cosmology, the different conversion of radial and transverse distances would lead to apparent anisotropies in the observed galaxy clustering, independently of RSDs.

As a result, any wave number~$k$ and angle variable~$\mu$ measured in the fiducial cosmology are in fact rescaled from the true cosmological ones given by
\begin{subequations}
    \label{eq:AP-rescaled wave number and angle variable}
    \begin{align}
        k_\true &= k \qty[\alpha_\perp^{-2} + \qty(\alpha_\parallel^{-2} - \alpha_\perp^{-2}) \mu^2]^{1/2} \,, \\
        \mu_\true &= \frac{\mu}{\alpha_\parallel} \qty[\alpha_\perp^{-2} + \qty(\alpha_\parallel^{-2} - \alpha_\perp^{-2}) \mu^2]^{-1/2} \,,
    \end{align}
\end{subequations}
where $\alpha_\parallel(z) = {\breve{H}(z)}\big/{H(z)}$~and~$\alpha_\perp(z) = {D_\angular(z)}\big/{\breve{D}_\angular(z)}$ are the scaling factors,\footnote{Sometimes the rescaling factors are defined with respect to the sound horizon~$r_\sound$ at the drag epoch~$z_\drag$, since the BAO feature in galaxy clustering can help break the degeneracy between RSDs and the AP effect~\cite{Beutler_2016}.} and the breve~$\breve{\placedot}$ denotes a quantity evaluated in the fiducial cosmology. Therefore in the presence of the AP effect, the observed anisotropic power spectrum should be~\cite{Ballinger_1996}
    \begin{equation}
        P(k, \mu) = \alpha_\parallel^{-1} \alpha_\perp^{-2} P_\true\qty(k_\true(k, \mu), \mu_\true(k, \mu)) \,.
        \label{eq:AP-rescaled power spectrum}
    \end{equation}

Even though our discussion so far is limited to linear perturbation theory, $P(k, \mu)$ remains the key observable predicted by models of non-linear galaxy clustering, e.g.~in the Taruya--Nishimichi--Saito (TNS) model for non-linear RSDs~\cite{Taruya_2010}. For observations of a small patch of the sky, the line of sight to each galaxy does not change much and the global plane-parallel $P(k, \mu)$ can be directly measured --- this is the case for pencil-beam like surveys of the past with very narrow opening angles, or for cosmological $N$-body simulations where the `observer' can be placed arbitrarily far away from the simulation box. However, for future large galaxy surveys such as DESI and \textit{Euclid} covering almost a third of the sky (approximately \num{14000} and \num{15000} square degrees respectively~\cite{DESI_2016,Euclid_2011}), the observer is at the centre of the much wider cosmic volume being probed --- any efforts to directly measure $P(k, \mu)$ will in effect be forcing the inherent spherical geometry of the survey to align with a Cartesian coordinate system~\cite{Castorina_2018}. This apparent mismatch will be the focus of our discussion in section~\ref{sec:Cartesian analysis} when we introduce the estimators for measuring the anisotropic power spectrum.

\subsection{Observational window effects}
\label{subsec:window effects}

Although the true background galaxy number density is expected to be spatially homogeneous in real space, i.e.~$\nbar(\vr, z) = \nbar(z)$, it can be modulated by spatial variations in incomplete observations so that
    \begin{equation}
        \nbar(\vr, z) = W(\vr) \nbar(z) \,,
        \label{eq:observable background number density}
    \end{equation}
where we assume the \emph{survey window}~$W(\vr)$ to be separable into angular and radial components,
    \begin{equation}
        W(\vr) = M(\vur) \phi(r) \,.
        \label{eq:spatial variation}
    \end{equation}
Here the \emph{angular mask} function~$M(\vur)$ may take binary values if it is simply a veto mask with 
    \begin{equation}
        \fsky = \frac{1}{4\uppi} \int \dd[2]{\vur} M(\vur) \in [0, 1]
        \label{eq:sky fraction}
    \end{equation}
being the fraction of the sky observed, but more generally $M(\vur)$ can include completeness variations and $\fsky$ becomes an effective sky fraction. The dimensionless \emph{radial selection} function~$\phi(r)$ is normal{\is}ed to the total number of galaxies in the survey volume,
    \begin{equation}
        N = 4\uppi \fsky \int \dd{r} r^2 \phi(r) \nbar(z) \,.
        \label{eq:selection function normalisation}
    \end{equation}
The window function can be extended to include radial or angular-dependent weights, as discussed in subsequent sections. By eqs.~\eqref{eq:galaxy over-density field} and \eqref{eq:galaxy--matter co-evolution}, the real-space galaxy number density observed is then
    \begin{equation}
        n(\vr, z) = M(\vur) \phi(r) \nbar(z) \qty[1 + b_k(z) D(z) \den_{\matter,0}(\vr)] \,.
        \label{eq:real-space galaxy number density}
    \end{equation}

The presence of the survey window affects the shape of the power spectrum by filtering $P(\vk)$ through $W(\vk)$ in Fourier space~\cite{Peacock_1991,Feldman_1994},
    \begin{equation}
        \filter{P}(\vk) = \int \frac{\dd[3]{\vq}}{(2\uppi)^3} \abs{W(\vk - \vq)}^2 P(\vq) \,,
        \label{eq:Fourier-space window convolution}
    \end{equation}
where the tilde $\filter{\vdot}$ denotes a window-convolved quantity. Rather than attempting to deconvolve the survey window from power spectrum measurements~\cite[see e.g.][]{Percival_2005}, which is an inverse operation that can amplify noise in the data, it is more desirable to forward model the convolved power spectrum~$\filter{P}(\vk)$~\cite{Wilson_2016}.

In the next section, we will set out how galaxy clustering can be measured and analysed in practice with a windowed anisotropic power spectrum, albeit with the line of sight varying across the survey volume. For this purpose, it is equivalent and more convenient to consider the multipoles of $\filter{P}(k, \mu)$ with respect to the Legendre polynomial $\legendre_\ell(\mu)$,
    \begin{equation}
        \filter{P}_\ell(k) = \frac{2\ell + 1}{2} \int_{-1}^{1} \dd{\mu} \legendre_\ell(\mu) \filter{P}(k, \mu) \,,
    \label{eq:power spectrum multipoles}
    \end{equation}
which are non-vanishing in the Kaiser RSD model only for the monopole, quadrupole and hexadecapole with~$\ell = 0, 2, 4$.

\section{Cartesian power spectrum analysis}
\label{sec:Cartesian analysis}

The standard Cartesian power spectrum analysis of galaxy survey catalogues usually follows the Feldman--Kaiser--Peacock (FKP) approach, where a high-density synthetic catalogue which is random and unclustered is used to probe non-uniform survey geometry and selection effects~\cite{Feldman_1994}. Therefore we consider the weighted galaxy number density field
\begin{subequations}
    \begin{equation}
        F(\vs) = \frac{w(\vs)}{I^{1/2}} [n_\gal(\vs) - \alpha n_\syn(\vs)] \,,
        \label{eq:FKP weighted field}
    \end{equation}
where $w(\vs)$ is some weighting scheme,
    \begin{equation}
        I \equiv \int \dd[3]{\vs} w(\vs)^2 \nbar(\vs)^2
        \label{eq:FKP weighted field normalisation}
    \end{equation}
is the normal{\is}ation constant and
    \begin{equation}
        \alpha = \frac{\sum_{i=1}^{N_\gal} w(s_i)}{\sum_{i=1}^{N_\syn} w(s_i)} \ll 1
        \label{eq:FKP weighted number count ratio}
    \end{equation}
\end{subequations}
is the ratio of the weighted number counts. Here $N_\gal$ is the number of galaxies in the survey catalogue and $N_\syn$ ($\gg N_\gal$) is the number count for the synthetic catalogue.

In the following subsections, we will outline the construction of an estimator~$\est{P}_\ell(k)$ from the weighted field~$F(\vs)$ that allows for a varying line of sight as well as the procedure for window convolution of the power spectrum multipole model~$P_\ell(k)$.

\subsection{Multipole estimation in the local plane-parallel approximation}

For realistic survey geometries, the assumption of a fixed line of sight simply does not hold --- this poses the question as to whether the global plane-parallel power spectrum~$P(k, \mu)$, or equivalently its multipoles~$P_\ell(k)$, can actually be recovered from clustering measurements. One attempt to address this problem, introduced by Yamamoto et al.~\cite{Yamamoto_2006}, is to adopt the so-called \emph{local}, or \emph{pairwise}, \emph{plane-parallel approximation}, in which the power spectrum multipoles are estimated by
    \begin{equation}
        \est{P}_\ell(k) = (2\ell + 1) \int \frac{\dd[2]{\vuk}}{4\uppi} \int \dd[3]{\vs_1} \e^{\im\vk\vdot\vs_1} \int \dd[3]{\vs_2} \e^{-\im\vk\vdot\vs_2} \legendre_{\ell}\qty\big(\vuk \vdot \vus_\lambda) F(\vs_1) F(\vs_2) \,,
        \label{eq:Yamamoto multipole estimator}
    \end{equation}
as one considers a pairwise line of sight
    \begin{equation}
        \vs_\lambda = \lambda \vs_1 + (1 - \lambda) \vs_2 \,, \quad 0 \leqslant \lambda \leqslant 1
        \label{eq:pairwise line of sight}
    \end{equation}
for two galaxies located at positions $\vs_1$~and~$\vs_2$ that contribute to the clustering 2-point function measurements~\cite{Yamamoto_2006,Castorina_2018}. For any value of $\lambda \in [0, 1]$, $\vs_\lambda$ lies between $\vs_1$~and~$\vs_2$ and is considered equivalent, i.e.~locally parallel; however, $\vs_\lambda$ itself is non-local since it is depends on a pair of galaxy positions.

The original Yamamoto estimator for power spectrum multipoles is given by eq.~\eqref{eq:Yamamoto multipole estimator} with the choice of $\lambda = 1/2$, i.e.~the line of sight~$\vs_\lambda$ is the mid-point between a galaxy pair. In this case, the computation of the estimator~$\est{P}_\ell(k)$ is costly as nested integrals have to be evaluated in sequence. However, refs.~\cite{Bianchi_2015,Scoccimarro_2015} have shown that, if one adopts the end-point line of sight~$\vs_\lambda$ with $\lambda = 0$~or~$1$, then the expression above can be made amenable to FFTs by considering the Cartesian components of $\vk$ separately. Ref.~\cite{Hand_2017} has further pointed out that the computation can become more efficient if one instead performs FFTs of the weighted field~$F(\vs)$ with spherical harmonics~$Y_{\ell m}$,
    \begin{equation}
        F_\ell(\vk) = \frac{4\uppi}{2\ell + 1} \sum_m \sphYc{\ell m}(\vuk) \int \dd[3]{\vs} \e^{-\im\vk\vdot\vs} \sphYc{\ell m}(\vus) F(\vs) \,,
        \label{eq:spherical harmonic transformed weighted field}
    \end{equation}
and then evaluates the estimator as
    \begin{equation}
        \est{P}_\ell(k) = (2\ell + 1) \int \frac{\dd[2]{\vuk}}{4\uppi} \conj{F}_0(\vk) F_\ell(\vk) \,.
        \label{eq:FFT-based Yamamoto multipole estimator}
    \end{equation}

The FFT-based Yamamoto estimator holds two chief advantages: it is computationally fast and it allows for a varying line of sight across the survey volume, which is far more realistic than fixing a global line of sight. Despite the anisotropies induced around the line of sight in redshift space, galaxy clustering remains statistically isotropic around the observer, and thus one only needs to assume the line of sight is `locally' parallel between a pair of galaxies, which can be rotated together freely in the observer's coordinate system --- it is for this reason that the Yamamoto estimator $\est{P}_\ell(k)$ mostly recovers the theoretical observable~$P(k, \mu)$, or equivalently $P_\ell(k)$, albeit convolved with the survey window.

\subsection{Window convolution of the model}
\label{subsec:window convolution}

As discussed in section~\ref{subsec:window effects}, models of the galaxy power spectrum need to be convolved with the survey window in order to account for the effect of non-uniform survey geometry in clustering measurements. This perhaps can be more easily understood in configuration space, where convolution of the power spectrum in Fourier space becomes multiplication of the galaxy clustering correlation function~$\xi(\vDelta)$ with
    \begin{equation}
        Q(\vDelta) = \int \dd[3]{\vs} W(\vs + \vDelta) W(\vs) \,,
        \label{eq:window auto-correlation function}
    \end{equation}
i.e.~$\filter{\xi}(\vDelta) = Q(\vDelta) \xi(\vDelta)$, where $\vDelta$ is the separation vector between two galaxy positions. Here $Q(\vDelta)$ is the inverse Fourier transform of $\abs{W(\vk)}^2$; although omitted in the definition above, $Q(\vDelta)$ is in practice normal{\is}ed to $Q(0) = 1$ and thus dimensionless.

Based on this result, ref.~\cite{Wilson_2016} has suggested that an efficient method to compute the window-convolved $\filter{P}_\ell(k)$ model is via the configuration space: one simply needs to first obtain the correlation function multipoles from $P_\ell(k)$ by the inverse Hankel transform,
    \begin{equation}
        \xi_\ell(\varDelta) = 4\uppi \im^\ell \int \frac{\dd{k} k^2}{(2\uppi)^3} j_\ell(k\varDelta) P_\ell(k) \,;
    \end{equation}
then perform the matrix multiplication
    \begin{equation}
        \filter{\xi}_\ell(\varDelta) = \mat{Q}_{\ell \ell'}(\varDelta) \xi_{\ell'}(\varDelta)
        \label{eq:correlation multipole multiplication}
    \end{equation}
where $\mat{Q}_{\ell \ell'}$ are entries of the matrix
    \begin{equation}
        \mat{Q} = %
        \begingroup
        \renewcommand*{\arraystretch}{1.5}
        \setlength\arraycolsep{4pt}
        \begin{pmatrix*}[r]
            Q_0 & \frac{1}{5} Q_2 & \frac{1}{9} Q_4 \\
            Q_2 & Q_0 + \frac{2}{7} Q_2 + \frac{2}{7} Q_4 & \frac{2}{7} Q_2 + \frac{100}{693} Q_4 + \frac{25}{143} Q_6 \\
            Q_4 & \frac{18}{35} Q_2 + \frac{20}{77} Q_4 + \frac{45}{143} Q_6 & Q_0 + \frac{20}{77} Q_2 + \frac{162}{1001} Q_4 + \frac{20}{143} Q_6 + \frac{490}{2431} Q_8
        \end{pmatrix*}
        \endgroup
        \label{eq:window matrix}
    \end{equation}
consisting of the survey window multipoles~$Q_\ell(\varDelta)$;\footnote{Both the correlation function multipoles~$\xi_\ell$ and the survey window multipoles~$Q_\ell$ are defined in a similar way to eq.~\eqref{eq:power spectrum multipoles}, except here the argument of the Legendre polynomials is the angle between the line of sight~$\los$ and the separation vector~$\vDelta$.} and finally Hankel transform back to
    \begin{equation}
        \filter{P}_\ell(k) = 4 \uppi \im^{-\ell} \int \dd{\varDelta} \varDelta^2 j_\ell(k\varDelta) \filter{\xi}_\ell(\varDelta) \,.
        \label{eq:convolved power spectrum multipoles}
    \end{equation}

Despite the aforementioned advantages of the Yamamoto estimator~$\est{P}_\ell(k)$, it cannot match the window-convolved model~$\filter{P}_\ell(k)$ in entirety. This is evident from the fact that $\est{P}_\ell(k)$ still depends on the choice of $\lambda \in [0, 1]$, which is not unique; indeed, for the end-point line of sight chosen for FFT computations, the exchange symmetry between a pair of galaxies is broken and odd power spectrum multipoles become non-zero, which is not predicted by the Kaiser RSD model. Furthermore, the mode-coupling term neglected by the distant-observer approximation (eq.~\ref{eq:distant-observer approximation}) is still missing, and the angle variable~$\mu = \vuk \vdot \vus_\lambda$ as the replacement of $\mu = \vuk \vdot \los$ is in fact ill-defined in Fourier space, since formally the variable~$\vus_\lambda$ should have been integrated over in Fourier transform~\cite{Castorina_2018}.

The discrepancy between the estimator's expectation~$\big\langle{\est{P}_\ell(k)}\big\rangle$ and the global plane-parallel $\filter{P}_\ell(k)$ is commonly known as the \emph{wide-angle effect},\footnote{Strictly speaking, the term `wide-angle effect' refers to the discrepancy caused by the plane-parallel approximations, but some literature reserves this term for the mode-coupling term in the RSD operator that is neglected in the distant-observer approximation (see ref.~\cite{Yoo_2014} for more detailed clarification).} which has been shown to scale as $k^{-2}$. Hence it can be a significant systematic on very large scales in the Cartesian galaxy clustering analysis, affecting studies of primordial non-Gaussianity and relativistic effects~\cite{Yoo_2014,Beutler_2019}. Efforts have already been made to circumvent or correct for wide-angle effects present in the galaxy clustering 2-point correlators: in configuration space, the correlation function can be expanded using bi-polar or tri-polar spherical harmonics~\cite{Szalay_1998,Szapudi_2004,Papai_2008}, which are valid for wide angular separations and also account for the mode-coupling term in the RSD operator (eq.~\ref{eq:RSD operator}) that is usually ignored; in Fourier space, only recently have perturbative corrections to the power spectrum multipoles been derived by refs.~\cite{Reimberg_2016,Castorina_2018,Beutler_2019}, which are complete in linear perturbation theory and recover the plane-parallel limit when the clustering scale is much smaller than the distance to galaxy pairs contributing to the 2-point function.

However, at a more fundamental level, the power spectrum analysis of galaxy clustering performed in the Cartesian coordinate system is a mismatch to the inherently spherical geometry of LSS observations. The question remains as to whether one could avoid all the geometric approximations discussed above. Indeed, the \emph{spherical Fourier analysis}, also known as spherical harmonic analysis, is such an alternative description of anisotropic clustering --- in the next section, we shall review both the modelling and measurements of fluctuations in the galaxy distribution with spherically decomposed clustering statistics.

\section{Spherical Fourier analysis}
\label{sec:spherical analysis}

The notion of Fourier transform can be general{\is}ed to \emph{harmonic analysis}, which describes a field~$f(\vr)$ by its decomposition in an orthogonal basis of eigenfunctions of the Laplacian~$\nabla^2$. We saw in section~\ref{sec:anisotropic clustering} that the inherent geometry of galaxy survey observations is spherical with the observer placed at the origin, so it would be more natural to consider a Fourier analysis of clustering measurements in the spherical coordinate system with a basis different from the Cartesian plane waves~$\set{\e^{\im\vk\vdot\vr}}$. In the following subsections, we will introduce the discrete spherical Fourier--Bessel transform suited for a survey volume of finite size, and review the spherically decomposed clustering statistics first introduced in refs.~\cite{Fisher_1995,Heavens_1995} but with a few extensions to their model(s).

\subsection{Spherical Fourier--Bessel transform}

In spherical coordinates~$\vr = (r, \vur)$, the analogue to the usual Fourier transform for a cosmological field~$f(\vr) \in \R^3$ is given by
    \begin{equation}
        f_{\ell m}(k) = \int \dd[3]{\vr} j_\ell(kr) \sphYc{\ell m}(\vur) f(\vr) \,, \quad f(\vr) = \frac{2}{\uppi} \sum_{\ell m} \int \dd{k} k^2 j_\ell(kr) \sphY{\ell m}(\vur) f_{\ell m}(k) \,,
        \label{eq:SFB transform}
    \end{equation}
where $j_\ell$ is the spherical Bessel function of the first kind of order~$\ell$, and $\sphY{\ell m}$ is the spherical harmonic function of degree~$\ell$ and order~$m$, with $\ell \in \mathbbm{N}$~and~$m = 0, \pm 1, \dots, \pm \ell$.\footnote{Note that the normal{\is}ation convention of the spherical Fourier--Bessel transform here differs slightly from those in refs.~\cite{Fisher_1995,Heavens_1995,Shapiro_2012,Castorina_2018}; ours is more similar to the Cartesian Fourier transform where any normal{\is}ation constants are attached to the inverse transform. The spherical Fourier--Bessel bases defined in these works and ours also differ from those in refs.~\cite{Leistedt_2012,Yoo_2013,Pratten_2013,Lanusse_2015} where the orthonormality condition is defined with respect to an inner product integral with a 1-dimensional measure. In practice, all these definitions are equivalent.} If the Dirichlet boundary condition~$f(\vr) = 0$ is imposed on a sphere at radius~$r = R$, the wave numbers are then discret{\is}ed,
    \begin{equation}
        k_{\ell n} = \frac{u_{\ell n}}{R} \,,
        \label{eq:discrete radial wave numbers}
    \end{equation}
where $u_{\ell n}$ is the $n$-th positive root of the spherical Bessel function~$j_\ell$. In this work, we shall refer to the associated tuple~$(\ell, m, n)$ as the \emph{spherical degree}, \emph{order} and \emph{depth}.

Therefore akin to the discrete Fourier transform over a regular Cartesian grid with periodic boundary conditions, a field~$f(\vr)$ that vanishes outside some maximum radius $R$ can be expanded in the \emph{spherical Fourier--Bessel} (SFB) basis~$\set{\ket{\mu} = \ket{\ell_\mu, m_\mu, n_\mu}}$,
    \begin{equation}
        f(\vr) = \sum_{\mu} \kappa_\mu f_\mu \braket{\vr}{\mu} \,, \quad f_\mu = f_{\ell_\mu m_\mu}(k_{\ell_\mu n_\mu})
        \label{eq:discrete SFB expansion}
    \end{equation}
where the normal{\is}ation coefficient $\kappa_\mu \equiv \kappa_{\ell_\mu n_\mu}$ is given by 
    \begin{equation}
        \kappa_{\ell n} = \frac{2}{R^3} j_{\ell+1}(u_{\ell n})^{-2} \,,
        \label{eq:SFB normalisation coefficient}
    \end{equation}
and $\braket{\vr}{\mu} = j_\mu(r) \sphY{\mu}(\vur)$ with~$j_\mu(r) \equiv j_{\ell_\mu}(k_{\ell_\mu n_\mu}r)$, $\sphY{\mu}(\vur) \equiv \sphY{\ell_\mu m_\mu}(\vur)$. In this basis, the Dirac and Kronecker delta functions are respectively represented by
    \begin{equation}
        \dirac(\vr - \vr') = \sum_{\mu} \kappa_\mu \braket{\vr}{\mu} \braket*{\mu}{\vr'} \,, \quad \kron_{\mu\nu} = \kappa_{\mu} \int_{\abs{\vr} < R} \dd[3]{\vr} \braket{\mu}{\vr} \braket{\vr}{\nu} \,.
        \label{eq:SFB delta function representations}
    \end{equation}

\subsection{Spherical decomposition of clustering statistics}
\label{subsec:spherical decomposition}

Since the SFB transform does not require the calculation of transverse distances from the angular position of galaxies, we can model the RSD and AP effects jointly by considering the redshift-space radial coordinate
    \begin{equation}
        s = \breve{\chi}(z^\textrm{obs}) = \breve{r} + \gamma u \,,
        \label{eq:redshift-space fiducial radial coordinate mapping}
    \end{equation}
where the breve $\breve{\placedot}$ denotes a fiducial distance--redshift relationship~$\breve{r} = \breve{\chi}(z)$, and the function
    \begin{equation}
        \gamma(z) \equiv \dv{\breve{\chi}}{\chi} = \frac{H}{\mathrm{c}} \dv{\breve{\chi}}{z}
        \label{eq:differential AP rescaling function}
    \end{equation}
measures the rescaling of the differential comoving distance from the true cosmology~\cite{Nicola_2014}.

Similar to the usual RSD derivation, we first use the local galaxy number conservation law~$n(\vr) \dd[3]{\vr} = n(\vs) \dd[3]{\vs}$ and eq.~\eqref{eq:real-space galaxy number density} to write the SFB coefficient of the redshift-space galaxy number density~$n(\vs)$ as
    \begin{align}
        n_\mu &= \int \dd[3]{\vs} j_\mu(s) \sphYc{\mu}(\vus) w(s) n(\vs) \nonumber \\
        &= \int \dd[3]{\vr} w(s) j_\mu(s) \sphYc{\mu}(\vur) M(\vur) \phi(r) \nbar(z) \qty[1 + b_k(z) D(z) \den_{\matter,0}(\vr)] \,,
        \label{eq:number density SFB transform}
    \end{align}
where $w(s)$ is the radial weighting scheme as before, and redshift-dependent quantities are implicitly integrated via the distance--redshift relationship. Note that for a galaxy survey with maximum redshift~$z_\textrm{max}$, the observed galaxy number density field vanishes outside $R = \breve{\chi}(z_\textrm{max})$ and thus the wave numbers are discrete. Using eq.~\eqref{eq:redshift-space fiducial radial coordinate mapping}, we can expand
    \begin{equation}
        w(s) \simeq w(\breve{r}) + \gamma u w'(\breve{r}) \,, \quad j_\mu(s) \simeq j_\mu(\breve{r}) + \gamma u j'_\mu(\breve{r}) \,,
        \label{eq:linear RSD expansions}
    \end{equation}
where a prime $'$ denotes the derivative of a function with respect to its argument. Note that these expansions up to the first derivative terms only are complete in linear perturbation theory, because the expansion parameter is proportional to $u$. By considering the SFB~expansion
    \begin{equation}
        \den_{\matter,0}(\vr) = \sum_\mu \kappa_\mu j_\mu(r) \sphY{\mu}(\vu{r}) \qty(\den_{\matter,0})_\mu
        \label{eq:matter density SFB expansion}
    \end{equation}
as well as the linear continuity equation
    \begin{equation}
        u = - f \vur \vdot \grad \nabla^{-2} \den_\matter
        \label{eq:linear continuity equation}
    \end{equation}
where $f(z) = \dv*{\ln D(z)}{\ln a(z)}$ is the linear growth rate,\footnote{On large scales, velocity bias between galaxies and dark matter is negligible.} we can recast
    \begin{align}
        n_\mu = & \int \dd[3]{\vr} w(\breve{r}) j_\mu(\breve{r}) \sphYc{\mu}(\vu{r}) M(\vu{r}) \phi(r) \nbar(z) \nonumber \\
        & + \int \dd[3]{\vr} w(\breve{r}) j_\mu(\breve{r}) \sphYc{\mu}(\vu{r}) M(\vu{r}) \phi(r) \nbar(z) D(z) \sum_\nu \kappa_\nu j_\nu(r) \sphY{\nu}(\vu{r}) b_{k_\nu}(z) \qty(\den_{\matter,0})_\nu \nonumber \\
        & + \int \dd[3]{\vr} (wj_\mu)'(\breve{r}) \sphYc{\mu}(\vu{r}) M(\vu{r}) \phi(r) \nbar(z) \gamma(z) f(z) D(z) \sum_\nu \frac{\kappa_\nu}{k_\nu^2} j'_\nu(r) \sphY{\nu}(\vu{r}) \qty(\den_{\matter,0})_\nu \,.
        \label{eq:number density SFB coefficient contributions}
    \end{align}
This is a sum of three contributions, namely the background piece, the fluctuation piece and the RSD piece. By introducing the angular, radial and RSD coupling coefficients~\cite{Fisher_1995,Heavens_1995,Shapiro_2012}
\begin{subequations}
    \label{eq:spherical couplings}
    \begin{align}
        M_{\mu\nu} &= \int \dd[2]{\vu{r}} \sphYc{\mu}(\vu{r}) M(\vu{r}) \sphY{\nu}(\vu{r}) \,,
        \label{eq:angular couplings} \\
        \varPhi_{\mu\nu} &= \kappa_\nu \int \dd{r} r^2 w(\breve{r}) j_\mu(\breve{r}) j_\nu(r) \phi(r) \nbar(z) \frac{b_{k_\nu}(z)}{b_{k_\nu}(0)} D(z) \,,
        \label{eq:radial couplings} \\
        \varUpsilon_{\mu\nu} &= \frac{\kappa_\nu}{k_\nu^2} \int \dd{r} r^2 (wj_\mu)'(\breve{r}) j'_\nu(r) \phi(r) \nbar(z) \gamma(z) \frac{f(z)}{f(0)} D(z) \,,
        \label{eq:RSD couplings}
    \end{align}
\end{subequations}
we obtain a more compact expression
    \begin{equation}
        n_\mu = \nbar_\mu + \sum_\nu M_{\mu\nu} \varPhi_{\mu\nu} \qty(b_0)_\nu \qty(\den_{\matter,0})_\nu + \sum_\nu M_{\mu\nu} \varUpsilon_{\mu\nu} f_0 \qty(\den_{\matter,0})_\nu \,,
        \label{eq:number density SFB coefficient}
    \end{equation}
where $(b_0)_\mu \equiv b_{k_\mu}(0)$, $f_0 \equiv f(0)$ and
    \begin{equation}
        \nbar_\mu = \int \dd[3]{\vr} w(\breve{r}) j_\mu(\breve{r}) \sphYc{\mu}(\vu{r}) M(\vu{r}) \phi(r) \nbar(z) \,.
        \label{eq:background number density SFB coefficient}
    \end{equation}
    
This shows that redshift-space galaxy clustering statistics in linear perturbation theory can be decomposed into spherical clustering modes
    \begin{equation}
        D_\mu \equiv n_\mu - \nbar_\mu = \sum_\nu M_{\mu\nu} \qty[(b_0)_\nu \varPhi_{\mu\nu} + f_0 \varUpsilon_{\mu\nu}] \qty(\den_{\matter,0})_\nu \,,
        \label{eq:SFB clustering modes}
    \end{equation}
which satisfy $\ev{D_\mu} = 0$ with the 2-point function
    \begin{equation}
        \ev{D_\mu \conj{D}_\nu} = \sum_\lambda M_{\mu\lambda} \conj{M}_{\nu\lambda} \qty[(b_0)_\lambda \varPhi_{\mu\lambda} + f_0 \varUpsilon_{\mu\lambda}] \qty[(b_0)_\lambda \varPhi_{\nu\lambda} + f_0 \varUpsilon_{\nu\lambda}] \kappa_\lambda^{-1} P_{\matter,0}(k_\lambda) + \ev{\epsilon_\mu \conj{\epsilon}_\nu} \,.
        \label{eq:spherical 2-point function}
    \end{equation}
Here $P_{\matter,0}$ is the matter power spectrum at the current epoch $z = 0$, and the additional term
    \begin{equation}
        \ev{\epsilon_\mu \conj{\epsilon}_\nu} = M_{\mu\nu} \int \dd{r} r^2 j_\mu(\breve{r}) j_\nu(\breve{r}) w(\breve{r})^2 \phi(r) \nbar(z)
        \label{eq:shot noise 2-point function}
    \end{equation}
accounts for the shot noise contribution~\cite{Fisher_1995,Heavens_1995,Nicola_2014}.

It is worth commenting here that the SFB mode~$D_\mu$ is constructed from fluctuations in the galaxy number density field directly, and it is in general \emph{not} equivalent to the SFB coefficient of the over-density field --- this is because the latter is defined as a fraction of the observable background number density which may be spatially varying. For the underlying matter density contrast in real space, however, this is not an issue since it is statistically~homogeneous.

\subsection{Comparisons with the Cartesian power spectrum analysis}
\label{subsec:comparison and connections between analyses}

To motivate the hybrid-basis approach to galaxy clustering analysis in the next section, it is worth making connections as well as comparisons between clustering modes $D_\mu$~and~$\den(\vk)$ and the Fourier analyses based on the SFB modes and Cartesian power spectrum.

First of all, we note that $D_\mu$~and~$\den(\vk)$ both depend on galaxy biasing with respect to the underlying matter distribution and capture anisotropic clustering due to the RSD and AP effects on linear scales. No geometric approximations are required to obtain $D_\mu$, and the spherical coupling coefficients $\varUpsilon_{\mu\nu}$ explicitly mix clustering modes at different wave numbers as a reflection of RSDs; in contrast, $\den(\vk)$ obtained in the distant-observer limit ignores the mode-coupling term in the Jacobian of the mapping from real to redshift space (see eq.~\ref{eq:RSD operator}). However, our derivation of the model of the SFB modes $D_\mu$ is limited to linear perturbation theory, and the extension to non-linear scales, though possible, is considerably more complex than models of the non-linear Cartesian power spectrum.

Secondly, the clear distinction between angular and radial components in the spherical Fourier analysis offers a number of advantages:
\begin{itemize}
    \item There is no ambiguity in the definition of the line of sight, which is free to vary across the entire sky, so wide-angle corrections are not needed;
    \item Relativistic and light-cone effects, which affect clustering along the line of sight, can be more easily included~\cite{Yoo_2013}. Indeed, redshift evolution in the galaxy population number density, clustering amplitude, galaxy bias and the growth rate is fully captured by the spherical couplings $\varPhi$~and~$\varUpsilon$ (eqs.~\ref{eq:radial couplings} and \ref{eq:RSD couplings}), where redshift dependence is integrated radially via the distance--redshift relationship;
    \item It allows easier joint analyses with other probes such as CMB, weak lensing and the integrated Sachs--Wolfe effect~\cite[e.g.][]{Webster_1998,Heavens_2003,Castro_2005,Shapiro_2012};
    \item The separation of angular and radial survey systematics allows individual clustering modes to be treated in isolation in analysis~\cite{Kitching_2014,Castorina_2018}.
\end{itemize}

Thirdly, the Cartesian power spectrum analysis is tomographic, i.e.~it requires binning and averaging in redshift, with models of $P_\ell(k)$ evaluated at some effective redshift in each bin. If the redshift bins are too narrow, large clustering modes along the line-of-sight can be missed and the level of shot noise is much higher; too wide, then the effective redshift cannot capture any redshift evolution within the bin. Indeed, the Fourier analysis based on $D_\mu$ has been shown be more robust and optimal in this regard~\cite[e.g.][]{Kitching_2014,Lanusse_2015}.

Lastly, the spherical Fourier analysis is based on individual modes~$D_\mu$, essentially a 1-point function of the cosmological field, whereas the Cartesian power spectrum analysis is based on the 2-point function compressed from many clustering modes. This is an important distinction with several implications:
\begin{itemize}
    \item In the case of the former, the distribution of SFB modes~$D_\mu$ is \emph{exactly} Gaussian as long as cosmic fluctuations can be described by Gaussian random fields. Although cosmic fields with non-zero PNG are not exactly Gaussian, the deviation is constrained to be small and any signature of PNG will be reflected more in the amplitude of clustering statistics rather than the overall probability distribution. The Gaussian random field assumption still serves as a useful null hypothesis for detecting any non-Gaussianity. In this case, all the cosmological information is encoded in the covariance matrix which is a 2-point function and analytically tractable. However, without the benefit of FFT algorithms, the computational cost of evaluating the 2-point function~\eqref{eq:spherical 2-point function} can be considerable, not least because angular integration over spherical harmonics and radial integration over spherical Bessel functions need to be performed repeatedly for all the coupling coefficients in eq.~\eqref{eq:spherical couplings} for different cosmological models. In appendix~\ref{app:computational complexity}, we provided a more detailed account of the computational complexity of the spherical Fourier analysis;
    \item In contrast, the distribution of measured power spectrum multipoles is only \emph{approximately} Gaussian when the number of clustering modes is large so the central limit theorem holds. The power spectrum models as well as measurements already compressed from individual Cartesian clustering modes can be efficiently computed using FFTs and the Hankel transform. However, the covariance matrix is now a 4-point function, which usually has to be estimated from ideally order ${\numrange[range-phrase=\textrm{--}]{e3}{e4}}$ realistic mock catalogues and poses a significant computational challenge~\cite{Hartlap_2006,Dodelson_2013,Taylor_2013,Percival_2014,Sellentin_2015}, though recently there have been some notable progress in obtaining analytic covariance matrices~\cite[e.g.][]{Li_2019,Wadekar_2019}.
\end{itemize}

Despite some of these apparent differences between $D_\mu$ and $\den(\vk)$ (or the power spectrum), in the simplest scenarios there are straightforward connections between the two thanks to the orthogonality of the SFB and plane wave bases. If the angular mask function $M(\vur) \equiv 1$, then the angular coupling coefficients~\eqref{eq:angular couplings} reduce to $M_{\mu\nu} = \kron_{\ell_\mu \ell_\nu} \kron_{m_\mu m_\nu}$; if in addition, there are no radial selection, weighting, AP effects or redshift evolution (i.e.~a fixed redshift is considered), then $M_{\mu\nu} \varPhi_{\mu\nu} = \kron_{\mu\nu}$. Therefore, in the absence of RSDs (i.e.~$\varUpsilon_{\mu\nu} \equiv 0$), the spherical 2-point function reduces from an infinite series~\eqref{eq:spherical 2-point function} to being diagonal,
    \begin{equation}
        \frac{\kappa_{\mu\nu}}{\nbar^2} \ev{D_\mu \conj{D}_\nu} = \kron_{\mu\nu} \qty[b_{k_\mu}^2 P_\matter(k_\mu) + \frac{1}{\nbar}] \,,
        \label{eq:2-point function diagonal reduction}
    \end{equation}
where we recogn{\is}e the right-hand side as simply the isotropic galaxy power spectrum plus the Poissonian shot noise. This also hints at a convergence check for the integrals~\eqref{eq:spherical couplings} and the series~\eqref{eq:spherical 2-point function}. Finally, for the sake of completeness, it is worth commenting here that under less restrictive assumptions than the above, the spherical 2-point function can also be related to the tomographic angular power spectrum~$C_\ell$, but we refer the reader to e.g.~refs.~\cite{Yoo_2013,Lanusse_2015,Castorina_2018} for more detail.

\section{Hybrid-basis likelihood inference}
\label{sec:hybrid basis approach}

Having laid out the different aspects of Fourier analyses based on the SFB and Cartesian plane wave bases, we propose a hybrid-basis approach to cosmological parameter inference from galaxy clustering measurements:
\begin{itemize}
    \item Since the survey geometry and other observational systematics have the biggest impact on the relatively few Fourier modes on the largest scales, we use the spherical Fourier analysis to faithfully capture the physics of anisotropic galaxy clustering in linear perturbation theory, and construct from the SFB modes $D_\mu$ the cosmological likelihood directly, which is exactly multivariate normal provided cosmic fluctuations are well described by Gaussian random fields. In this case, the covariance matrix is the 2-point function, which contains all the cosmological information and can be computed~analytically;
    \item On comparatively smaller scales, we choose the Yamamoto estimator~$\est{P}_\ell(k)$ for power spectrum multipoles as our summary statistics, which can be efficiently computed using FFTs. Since $\est{P}_\ell(k)$ is compressed from a large number of Cartesian Fourier modes, its probability distribution is very close to being Gaussian by the central limit theorem;
    \item By combining the probability distributions of $D_\mu$~and~$\est{P}_\ell(k)$, we can then obtain a hybrid-basis likelihood for cosmological parameter $\theta$.
\end{itemize}

This idea of adopting different statistics, either uncompressed or compressed from individual modes of fluctuations depending on the physical scale considered, is inspired by the use of hybrid estimators in CMB studies~\cite{Efstathiou_2004,Hinshaw_2007}: one approach is to evaluate the likelihood from the CMB map pixels directly, either searching for a quadratic or maximum likelihood estimator of the angular power spectrum~\cite{Tegmark_1997a,Bond_1998,Bond_2000} or Monte Carlo sampling the posterior surface~\cite{Jewell_2004,Wandelt_2004,Eriksen_2004}; another is to compress the map pixels into pseudo-$C_\ell$ estimators of the angular power spectra based on which an approximate likelihood can be constructed~\cite{Wandelt_2001,Hamimeche_2008}. This strategy has then been successfully applied by \textit{Planck} to its cosmological likelihoods, which consist of a low-$\ell$ part ($\ell \leqslant 29$) based on the CMB temperature and polar{\is}ation map pixels, and a high-$\ell$ part ($\ell \geqslant 30$) based on the pseudo-$C_\ell$ estimator~\cite{Planck_2014,Planck_2016,Planck_2019}. As far as we are aware, this approach has not been applied in any LSS settings before, which are arguably more nuanced as LSS data sets are intrinsically 3-dimensional. We will set out in this section the basic steps involved in constructing the likelihood functions~$\Like{\theta}$ for $D_\mu$~and~$\est{P}_\ell(k)$, which are split at the \emph{hybrid{\is}ation scale} $k_\hyb$ analogous to the $\ell$ split in \textit{Planck} likelihoods, i.e.~we restrict the SFB wave numbers to $k_\mu \leqslant k_\hyb$ and the power spectrum wave numbers to $k_\hyb < k < k_\max$, where $k_\max$ is the overall maximum wave number in the analysis.

\paragraph{Spherical-basis likelihood.} The data vector of SFB modes, $\vb{D} \equiv (D_\mu)$, is calculated from the survey and synthetic catalogues by direct summation over weighted delta function contributions from each galaxy (see eq.~\ref{eq:SFB clustering modes}),
    \begin{equation}
        D_\mu = \sum_{i=1}^{N_\gal} w(s_i) j_\mu(s_i) \sphYc{\mu}(\vus_i) - \alpha \sum_{i=1}^{N_\syn} w(s_i) j_\mu(s_i) \sphYc{\mu}(\vus_i) \,.
        \label{eq:SFB direct summation}
    \end{equation}
Here the vector index~$\mu$ can either be `naturally' ordered by the tuple~$(\ell_\mu, m_\mu, n_\mu)$, or `spectrally' ordered by the wave number~$k_\mu$ and the spherical order~$m_\mu$. Since fluctuations in the galaxy distribution are well described by a Gaussian random field on large scales and the SFB transform is linear in the field, the spherical-basis data vector follows the \emph{circularly-symmetric complex normal distribution}, $\vb{D} \sim \ComplexNormal{\vb{0}, \mat{C}}$.\footnote{In contrast to previous works~\cite[e.g.][]{Heavens_1995}, we do not separate the spherical-basis data vector into real and imaginary parts which jointly follow the multivariate normal distribution.} Therefore the spherical-basis likelihood function is given by the probability density function (PDF)~\cite{Wooding_1956,Goodman_1963}
    \begin{equation}
        \Like[\sph]{\theta} = \Prob[\big]{\given[\big]{\vb{D}}{\theta}} = \frac{\exp\!\qty\big[- \herm{\vb{D}} \mat{C}(\theta)^{-1} \vb{D}]}{\abs{\uppi\mat{C}(\theta)}} \,,
        \label{eq:spherical-basis likelihood}
    \end{equation}
where all cosmological parameter dependence is in the covariance matrix~$\mat{C}(\theta) = \Cov[\big]{\vb{D}}$ whose entries~$\mat{C}_{\mu\nu} = \ev{D_\mu \conj{D}_\nu}$ are precisely the spherical 2-point function~\eqref{eq:spherical 2-point function}. However, we would like to point out a few technicalities in the practical evaluation of $\Like[\sph]{\theta}$ above:
\begin{itemize}
    \item Based on the symmetry of the spherical harmonics, $\sphYc{\ell{\mkern 2mu}-m}(\vus) = (-1)^m \sphY{\ell m}(\vus)$, almost half of the SFB modes~$D_\mu$ can be calculated simply using $D_{\ell{\mkern 2mu}-m n} = (-1)^m \conj{D}_{\ell m n}$;
    \item In order to evaluate $\mat{C}_{\mu\nu} = \ev{D_\mu \conj{D}_\nu}$ accurately, we must ensure the infinite series~\eqref{eq:spherical 2-point function} numerically converges. This requires additional modes with wave numbers~$k_\lambda > k_\hyb$ to be included in the sum, and the appropriate truncation point in the series may need to be determined empirically, e.g.~using the diagonal 2-point function~$\big\langle{\abs{D_\mu}^2}\big\rangle$ (see eq.~\ref{eq:2-point function diagonal reduction}) as a diagnostic quantity;
    \item Since the data vector~$\vb{D} \in \C^{N_\data}$ consists of $N_\data$ uncompressed SFB modes, dimensions of the covariance matrix~$\mat{C} \in \C^{N_\data \times N_\data}$ can be large enough that the inversion of $\mat{C}$ becomes numerically unstable when it is not diagonal, i.e.~when the SFB modes are correlated. In this case, some eigenvalues of $\mat{C}$ can be very close zero, and with imperfect numerical precision the inverted matrix~$\mat{C}^{-1}$ may acquire large negative eigenvalues, posing a significant challenge to the sampling of the posterior distribution from $\Like[\sph]{\theta}$. One possible remedy to this problem is to apply a compression matrix~$\mat{R} \in \C^{N'_\data \times N_\data}$ to both the data vector and the covariance matrix before evaluating $\Like[\sph]{\theta}$, i.e.~we replace
        \begin{equation}
            \vb{D} \mapsto \mat{R} \vb{D} \,, \quad \mat{C} \mapsto \mat{R} \mat{C} \trans{\mat{R}}
            \label{eq:spherical-basis data compression}
        \end{equation}
    where $N'_\data < N_\data$~and~$\mat{R} \trans{\mat{R}} = \mat{I}$ is the identity matrix. In appendix~\ref{app:data compression}, we discuss one such compression method to ensure numerical stability.
\end{itemize}

\paragraph{Cartesian-basis likelihood.} To construct the weighted field~$F(\vs)$ from the survey and synthetic catalogues (eq.~\ref{eq:FKP weighted field}), the galaxy number density fields $n_\gal(\vs)$~and~$n_\syn(\vs)$ need to be interpolated on a regular Cartesian grid. The transformed quantity~$F_\ell(\vk)$ (eq.~\ref{eq:spherical harmonic transformed weighted field}) should be then be compensated for the interpolation kernel after FFTs~\cite{Jing_2005}. We denote the data vector of the estimated power spectrum multipoles by $\est{\vb{P}} = \qty\big(\est{P}_{\ell}(k_i))$, where the components are ordered by the multipole order $\ell$ and then the wave number bin $i$. Being the 2-point function of a Gaussian random field, the data vector~$\est{\vb{P}}$ should follow the hypo-exponential distribution~\cite{Wang_2019}; however, in the central limit theorem when the number of clustering modes contributing to $\est{P}_\ell(k_i)$ is large, one could assume the multivariate normal distribution~$\est{\vb{P}} \sim \Normal[\big]{\widebar{\vb{P}}, \mat{\Sigma}}$. The Cartesian-basis power spectrum likelihood function is thus given by
    \begin{equation}
        \Like[\Cart]{\theta} = \Prob[\big]{\given[\big]{\est{\vb{P}}}{\theta}} = \abs{2\uppi\mat{\Sigma}(\theta)}^{-1/2} \exp{- \frac{1}{2} \trans{\qty\Big[\est{\vb{P}} - \widebar{\vb{P}}(\theta)]} \mat{\Sigma}(\theta)^{-1} \qty\Big[\est{\vb{P}} - \widebar{\vb{P}}(\theta)]} \,.
        \label{eq:Cartesian-basis likelihood}
    \end{equation}
Here $\mat{\Sigma}(\theta) = \Cov[\big]{\est{\vb{P}}}$ is the covariance matrix, and $\widebar{\vb{P}}(\theta) = \Exp[\big]{\est{\vb{P}}}$ is the expectation of the power spectrum multipole estimator with components
    \begin{equation}
        \widebar{P}_\ell(k_i) = \filter{P}_\ell(k_i) + P_{\ell,\shot} \,,
        \label{eq:Cartesian-basis data expectation}
    \end{equation}
where $\filter{P}_\ell(k_i)$ is the window-convolved model of power spectrum multipoles (see section~\ref{subsec:window convolution}) and $P_{\ell,\shot}$ is the shot noise contribution~\cite{Bianchi_2015},
    \begin{equation}
        P_{\ell,\shot} = \frac{1 + \alpha}{I} \int \dd[3]{\vs} \legendre_\ell(\vuk\vdot\vus) w(s)^2 \nbar(\vs) \,.
        \label{eq:power spectrum multipole shot noise}
    \end{equation}
Similar to the spherical-basis likelihood, there are two technicalities related to the covariance matrix in evaluating the Cartesian-basis likelihood~$\Like[\Cart]{\theta}$:
    \begin{itemize}
        \item The true covariance matrix~$\mat{\Sigma}$ is usually analytically intractable, so it has to be replaced by an estimate~$\est{\mat{\Sigma}}$ from mock catalogues. Ref.~\cite{Sellentin_2015} has shown that the appropriate distribution to use as the likelihood function is no longer multivariate normal but a modified Student's distribution. However, when the number of mock catalogues used to obtain the estimate $\est{\mat{\Sigma}}$ far exceeds the dimension of the data vector~$\est{\vb{P}}$, the multivariate normal distribution remains an excellent approximation;
        \item Because of the high computational cost associated with generating a large number of mock catalogues, the covariance matrix estimate~$\est{\mat{\Sigma}}$ is usually produced at fixed fiducial cosmological parameters~$\theta_\fid$. To account for any parameter dependence, ref.~\cite{Wang_2019} proposed the variance--correlation decomposition which allows a parameter-dependent estimate~$\est{\mat{\Sigma}}(\theta)$ to be obtained from the fiducial estimate~$\est{\mat{\Sigma}}_\fid$ by a rescaling,
            \begin{equation}
                \est{\mat{\Sigma}}(\theta) = \mat{\Lambda}(\theta) \mat{\Lambda}_{\fid}^{-1} \est{\mat{\Sigma}}_\fid \mat{\Lambda}_{\fid}^{-1} \mat{\Lambda}(\theta) \,.
                \label{eq:variance--correlation rescaling}
            \end{equation}
        Here $\mat{\Lambda}(\theta) = \Diag[\big]{\widebar{P}_\ell(k_i)}$ is a diagonal matrix with entries given by the convolved power spectrum multipole model, including the shot noise contribution, at cosmological parameters~$\theta$, and $\mat{\Lambda}_\fid$ is the diagonal matrix evaluated at fiducial parameters~$\theta_\fid$.
    \end{itemize}

\paragraph{Hybrid-basis likelihood.} In the ideal{\is}ed scenario where the data vectors $\vb{D}$~and~$\est{\vb{P}}$ are independent, the hybrid-basis likelihood is simply the product of the two above, i.e.
    \begin{equation}
        \Like[\hyb]{\theta; \vb{D}, \est{\vb{P}}} = \Like[\sph]{\theta; \vb{D}} \Like[\Cart][\big]{\theta; \est{\vb{P}}} \,.
        \label{eq:hybrid-basis likelihood}
    \end{equation}
Unfortunately, this does not strictly hold when clustering modes of different wave numbers $k$ are mixed in the presence of survey window and RSD effects. In the \textit{Planck} likelihood analyses, the correlation between low-$\ell$ and high-$\ell$ components poses a similar issue, and different hybrid{\is}ation schemes were explored~\cite{Planck_2014,Planck_2016,Planck_2019}; they have found that the analysis results are not particularly sensitive to the hybrid{\is}ation scheme and thus a sharp transition between low-$\ell$ and high-$\ell$ components can be adopted without accounting for their correlation. In this work, we make a similar \emph{assumption} that the low-$k$ spherical and high-$k$ Cartesian likelihoods can be directly combined --- this is justified if their correlation is weak when the mixing kernel is sufficiently narrow in $k$-space and if the joint probability distribution of $\vb{D}$~and~$\est{\vb{P}}$ is multivariate normal.\footnote{Note that zero correlation does \emph{not} necessarily imply independence between two multivariate normal random variables \emph{unless} their joint probability distribution is also multivariate normal.} In practice, the correlation between the spherical-basis and Cartesian power spectrum data can be estimated from mock catalogues alongside the covariance matrix estimate for power spectrum multipoles, and one could attempt to decorrelate the combined data vector or reweight different data components before combining them, e.g.~with a Bayesian hyperparametric method~\cite{Lahav_2000,Hobson_2002,Ma_2014}.

In the next section, we will compare the Cartesian-basis power spectrum likelihood~\eqref{eq:Cartesian-basis likelihood} and the hybrid-basis likelihood~\eqref{eq:hybrid-basis likelihood} in a parameter inference problem to demonstrate the applicability of our new approach.

\section{Constraining primordial non-Gaussianity from simulations}
\label{sec:application}

It is now known that primordial non-Gaussianity, which encodes dynamics of the inflationary period in the early Universe, leaves an imprint in the late-time large-scale structure not only in higher-order statistics such as the bispectrum, but also in the clustering of virial{\is}ed haloes by introducing a scale-dependent modification to the tracer bias on large scales~\citep{Dalal_2008,Matarrese_2008,Slosar_2008}. In the presence of local PNG~$\fNL$, the linear galaxy bias~$b_1(z)$ receives a scale-dependent modification
    \begin{equation}
        \Delta b(k, z) = 3 \fNL (b_1 - p) \frac{1.3 \delta_\textrm{c} \varOmega_{\matter,0}}{k^2 T(k) D(z)} \left(\frac{H_0}{c}\right)^2 \,,
    \end{equation}
where $\delta_\textrm{c} \approx 1.686$ is the critical density of spherical collapse, $\varOmega_{\matter,0}$~and~$H_0$ are the matter density and Hubble parameters at the present epoch~$z = 0$, $c$ is the speed of light in vacuum, and $T(k)$ is the matter transfer function. Here we set the tracer-dependent parameter~$p = 1$,\footnote{The parameter~$p$ is usually set to $1$ for tracer samples selected by halo mass and $1.6$ for tracer samples dominated by recent halo mergers~\cite{Slosar_2008}. We have chosen $p = 1$ as it is a good match to our mock catalogues; however, ref.~\cite{Barreira_2020} has recently shown with simulations that $p$ even can be less than $1$ for haloes and galaxies selected by stellar mass.} and the numerical factor~$1.3$ arises as we normal{\is}e the linear growth factor~$D(z)$ to unity at present.\footnote{This normalisation factor actually depends on the value of~$\Omega_{\matter,0}$ which we specify later for our simulations.} The bias parameter~$b_k$ that appears in galaxy clustering statistics in the previous sections now includes this scale-dependent modification, i.e.~$b_k = b_1 + \Delta b$. As $k \rightarrow 0$, $T(k) \rightarrow 1$~and~$\Delta b \propto k^{-2}$, so the signature of $\fNL$ is enhanced. The sensitivity of $\fNL$ to large-scale clustering measurements makes it an ideal parameter to test our hybrid-basis approach to likelihood inference.

As a first step to demonstrate the applicability of the hybrid-basis approach, we employ halo mock catalogues generated from $N$-body simulations with (non-)Gaussian initial conditions and try to infer the local PNG parameter~$\fNL$ from the real-space halo clustering. In the next subsection~\ref{subsec:catalogue data}, we will describe the mock catalogue properties and the intermediary data products needed for likelihood evaluations, such as the survey window, covariance matrix estimates and the spherical coupling coefficients; in subsection~\ref{subsec:analysis results}, we compare parameter constraints on $\fNL$~and~$b_1$ from the hybrid-basis likelihood and the Cartesian-basis power spectrum likelihood.

\subsection{Mock catalogues and data products}
\label{subsec:catalogue data}

Our halo mock catalogues are generated from a series of dark matter $N$-body simulations at a flat {\textLambda}CDM cosmology with $(h, \varOmega_{m,0}, \varOmega_{b,0}, \sigma_8) = (0.70, 0.27, 0.044, 0.80)$. We first compute the matter transfer function with the public code \codename{camb}\footnote{Code for Anisotropies in the Microwave Background, \href{https://camb.info/}{\texttt{camb.info}}}~\cite{Lewis_2002}, which is then used to calculate initial conditions with the second-order Lagrangian perturbation theory~(2LPT). To seed the simulations, we make use of the public code \codename{2LPTic}\footnote{2LPT Initial Conditions, \href{https://cosmo.nyu.edu/roman/2LPT/}{\texttt{cosmo.nyu.edu/roman/2LPT}}} which can generate initial conditions with non-zero local PNG~\citep{Crocce_2006,Scoccimarro_2012}. In total, we have run 24~simulations with $\fNL = 0$ and 20~simulations with $\fNL = 100$, each in a $\SI{1}{\per\cubic\h\cubic\giga\parsec}$ comoving box of $512^3$ dark matter particles evolved from redshift $z = 32$~to~$z = 1$ using the public code \codename{gadget-2}\footnote{GAlaxies with Dark matter and Gas intEracT, \href{https://wwwmpa.mpa-garching.mpg.de/gadget/}{\texttt{mpa-garching.mpg.de/gadget}}}~\cite{Springel_2005}. We have not run the simulations down to redshift~$z = 0$ because of computation time; indeed, future galaxy surveys probing $\fNL$ on very large scales will mostly focus on the $z > 1$ Universe. Finally, we identify dark matter haloes within our mock catalogues using the public code \codename{ahf}\footnote{Adaptive Mesh Investigations of Galaxy Assembly (\codename{amiga}) Halo Finder, \href{http://popia.ft.uam.es/AHF}{\texttt{popia.ft.uam.es/AHF}}}~\cite{Knollmann_2009}, with at least $36$ particles per halo corresponding to a minimum halo mass of $M_\textrm{h} \approx \SI{2.0e13}{\per\h\solarmass}$.

\begin{figure}
    \centering
    \includegraphics[scale=0.775]{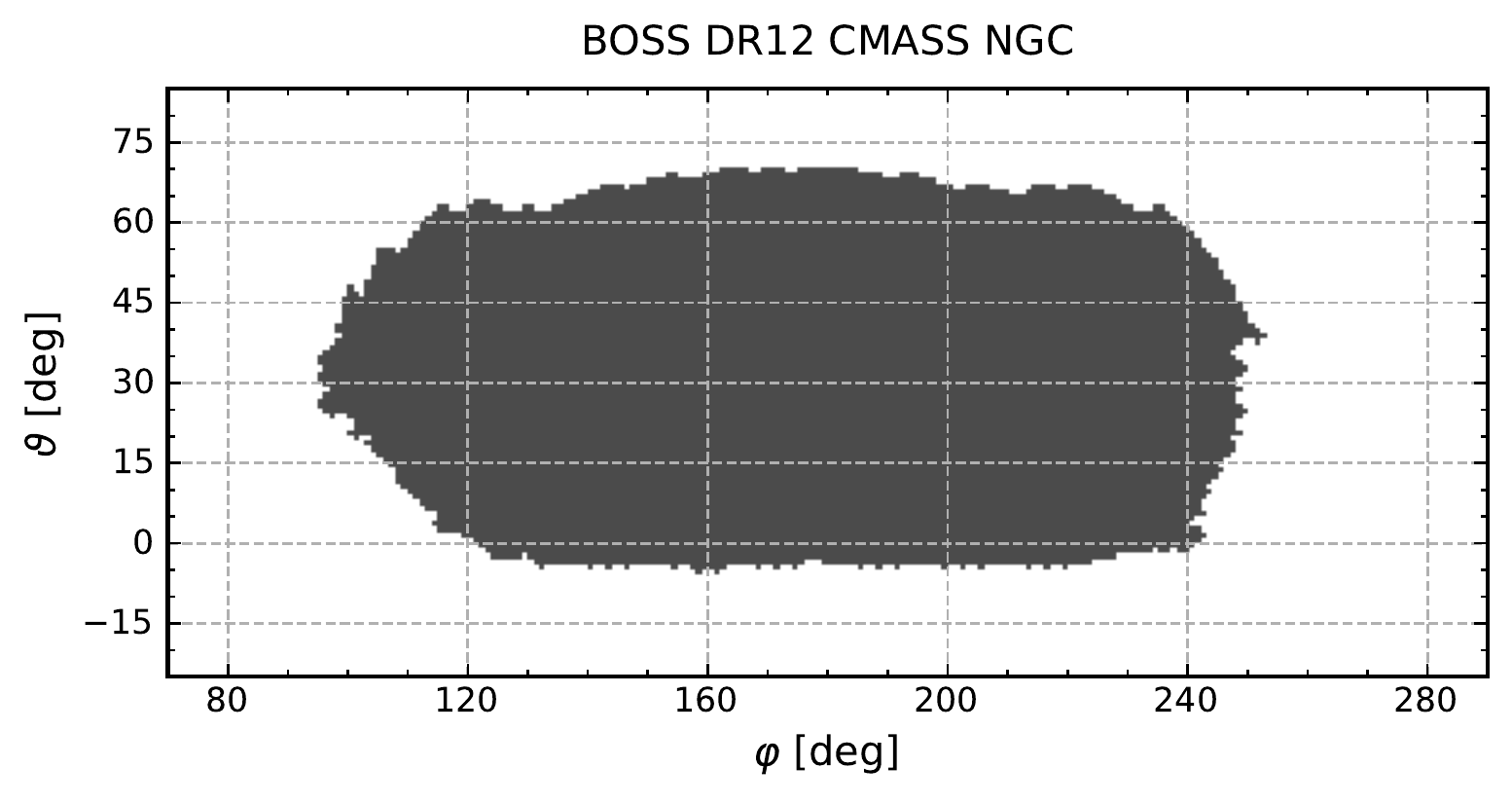}
    \caption{Survey angular mask $M(\vur)$ used in the partial-sky set-up. The angular mask function takes binary values ($1$ shown by the shaded region and $0$ elsewhere) and is constructed from the BOSS DR12 CMASS NGC random catalogue with \codename{HEALPix} pixelation~$N_\textrm{side} = 32$. The vertical and horizon axes correspond to the polar and azimuthal angles~$(\vartheta, \varphi)$ respectively in the spherical coordinate system.}
    \label{fig:survey mask}
\end{figure}

To compare the hybrid-basis and Cartesian power spectrum analyses, we consider two geometric set-ups: in the `full-sky' scenario, we only use the proportions of our mock catalogues within a comoving sphere of~radius $R = \SI{500}{\per\h\mega\parsec}$; in the `partial-sky' scenario, we further restrict the domain to the proportions covered by the footprint of the BOSS DR12 CMASS North Galactic Cap~(NGC) sample within the radius range~$\SI{100}{\per\h\mega\parsec} \leqslant r \leqslant R$. The angular mask function~$M(\vur)$ is constructed from the BOSS random catalogue\footnote{\href{https://data.sdss.org/sas/dr12/boss/lss/random0_DR12v5_CMASS_North.fits.gz}{\texttt{data.sdss.org/sas/dr12/boss/lss/random0\_DR12v5\_CMASS\_North.fits.gz}}} with \codename{HEALPix}\footnote{Hierarchical Equal Area isoLatitude Pixelization, \href{https://github.com/healpy/healpy}{\texttt{github.com/healpy/healpy}}} pixelation~$N_\textrm{side} = 32$ \cite{Gorski_2005}, as shown in figure~\ref{fig:survey mask}.
This angular mask corresponds to a sky fraction of $\fsky \approx 0.2$ and is simply chosen to demonstrate the hybrid-basis approach with a realistic survey geometry.

For both full-sky and partial-sky set-ups, we perform the hybrid-basis and Cartesian power spectrum analyses with maximum wave number~$k_\textrm{max} = \SI{0.08}{\h\per\mega\parsec}$.\footnote{Owing to the limited resolution of our simulations, we have found the halo bias to be slightly scale-dependent even in the absence of PNG. Therefore we have set a relatively high minimum halo mass in the mock catalogues and adopted $k_\max = \SI{0.08}{\per\h\mega\parsec}$ as the wave number upper cutoff.} As no anisotropies are expected from real-space halo clustering, we only consider the power spectrum monopole in wave number bins with uniform width~$\Delta k = \SI{0.01}{\h\per\mega\parsec}$. For the hybrid-basis analysis, we push the hybrid{\is}ation scale up to $k_\hyb = \SI{0.04}{\h\per\mega\parsec}$ for which the computation time of spherical clustering statistics and their likelihood function remains reasonable (see appendix~\ref{app:computational complexity}).

Before we evaluate the likelihood functions as outlined in section~\ref{sec:hybrid basis approach}, some intermediary data products in addition to the hybrid-basis and Cartesian power spectrum data vectors $\vb{D}$~and~$\est{\vb{P}}$ are required:
\begin{enumerate}
    \item the spherical coupling coefficients $M_{\mu\nu}$~and~$\varPhi_{\mu\nu}$, which are numerically computed from the angular mask $M(\vur)$ and the radial selection function $\phi(r)$ using eq.~\eqref{eq:spherical couplings};
    \item the survey window auto-correlation multipoles~$Q_\ell(\varDelta)$ used to convolve power spectrum models (see section~\ref{subsec:window convolution}), which can be determined from a synthetic random catalogue;
    \item the fiducial covariance matrix estimate $\est{\mat{\Sigma}}_\fid$ for the binned power spectrum monopole, which is obtained from a large number of synthetic random catalogues.
\end{enumerate}

\begin{figure}
    \centering
    \includegraphics[width=0.775\linewidth]{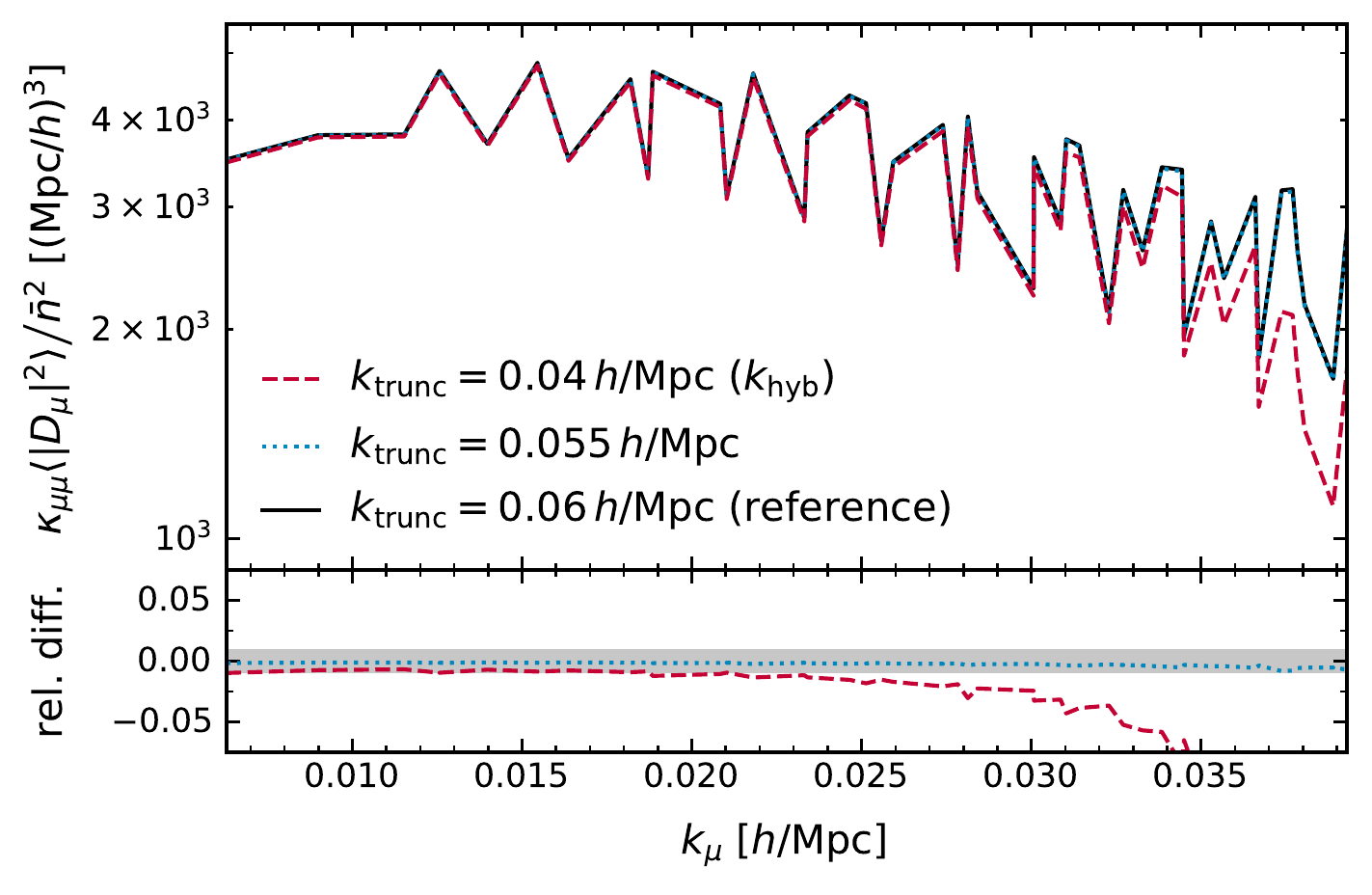}
    \caption{Convergence check of the normal{\is}ed diagonal spherical 2-point function, $\kappa_{\mu\mu} \big\langle\abs{D_\mu}^2\big\rangle\big/\nbar^2$, evaluated at each SFB mode wave number $k_\mu$ from the series~\eqref{eq:shot noise 2-point function}. The top panel shows the series truncated at different wave numbers~$k_\textrm{trunc} = 0.04, 0.055, 0.06\,\si{\per\h\mega\parsec}$ corresponding to $k_\hyb$ (dashed red line), the actual wave number cutoff adopted in our analysis (dotted blue line) and the reference case (solid black line). The bottom panel shows the relative difference of each series sum compared to the reference case, with the shaded region marking deviations within $\pm\SI{1}{\percent}$. Note that oscillations in the top panel are \emph{not} due to numerical noise but simply the behaviour of SFB modes.}
    \label{fig:convergence check}
\end{figure}

\begin{figure}
    \centering
    \includegraphics[width=.49\linewidth]{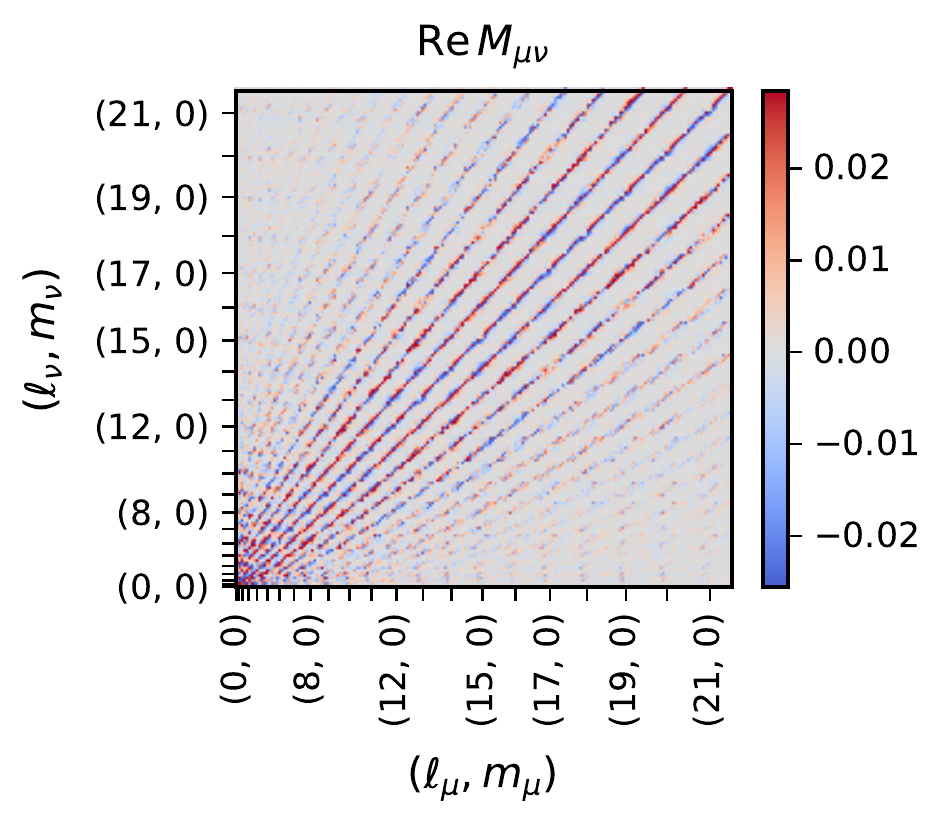}
    \hfill
    \includegraphics[width=.49\linewidth]{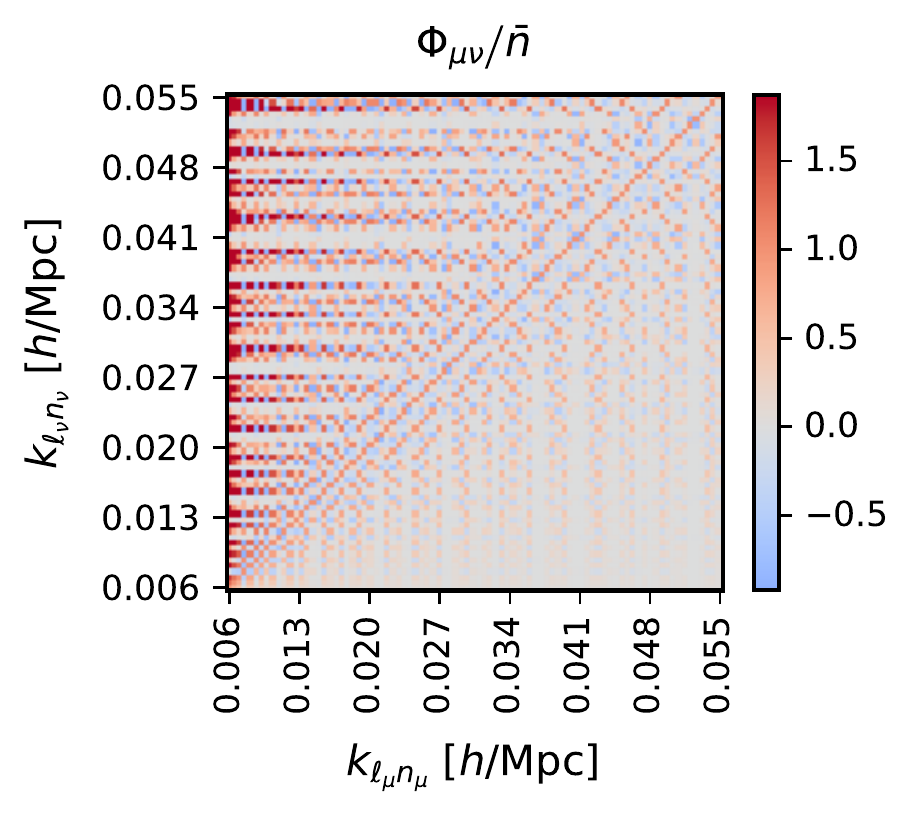}
    \caption{Matrices of the dimensionless angular and radial coupling coefficients $\operatorname{Re}{M_{\mu\nu}}$ (\textit{left column}) and $\varPhi_{\mu\nu}/\bar{n}$ (\textit{right column}) for wave numbers $k_{\ell n} \leqslant k_\textrm{trunc} = \SI{0.055}{\per\h\mega\parsec}$ in the partial-sky set-up. $M_{\mu\nu}$ coefficients are ordered by spherical degree and order~$(\ell, m)$ whereas $\varPhi_{\mu\nu}$ coefficients are ordered by wave number~$k_{\ell n}$.}
    \label{fig:spherical couplings}
\end{figure}

First, we consider the angular and radial spherical coupling coefficients $M_{\mu\nu}$~and~$\varPhi_{\mu\nu}$ required for computing the spherical 2-point function model. Although only SFB modes with wave numbers $k_\mu \leqslant k_\hyb$ are included in the data vector~$\vb{D}$, coupling between modes in the partial-sky case means that more modes with wave numbers~$k_\mu > k_\hyb$ must be included in the series~\eqref{eq:spherical 2-point function} for convergence. To check this, we use the normal{\is}ed diagonal 2-point function~$\kappa_{\mu\mu} \big\langle\abs{D_\mu}^2\big\rangle$ as a diagnostic quantity, and compare sums of the series truncated at wave numbers $k_\textrm{trunc} = 0.04, 0.055, 0.06\,\si{\per\h\mega\parsec}$,\footnote{We choose to evaluate the 2-point function model at our fiducial background cosmology with $\fNL = 0$ and $b_1 = 1$.} as shown in figure~\ref{fig:convergence check}. Note that oscillations shown in the figure are \emph{not} due to numerical noise; they are simply the behaviour of SFB modes.
We have found that $k_\textrm{trunc} = \SI{0.055}{\per\h\mega\parsec}$ is sufficient to ensure convergence at percent levels, which we adopt as the series cutoff in our spherical Fourier analysis. In figure~\ref{fig:spherical couplings}, we visual{\is}e matrices of the dimensionless coupling coefficients $\operatorname{Re}{M_{\mu\nu}}$~and~$\varPhi_{\mu\nu}/\nbar$ for wave numbers $k_\mu \leqslant k_\textrm{trunc} = \SI{0.055}{\per\h\mega\parsec}$. Entries of the angular coupling matrix~$\operatorname{Re}{M_{\mu\nu}}$ are arranged in the `natural' order by the $(\ell, m)$ tuple and entries of the radial coupling matrix~$\varPhi_{\mu\nu}$ are arranged in the `spectral' order by the wave number $k_\mu$. We do not show the imaginary part of $M_{\mu\nu}$ since it is close to zero for a binary-valued angular mask~\cite{Heavens_1995}.

\begin{figure}
    \centering
    \includegraphics[width=.49\linewidth]{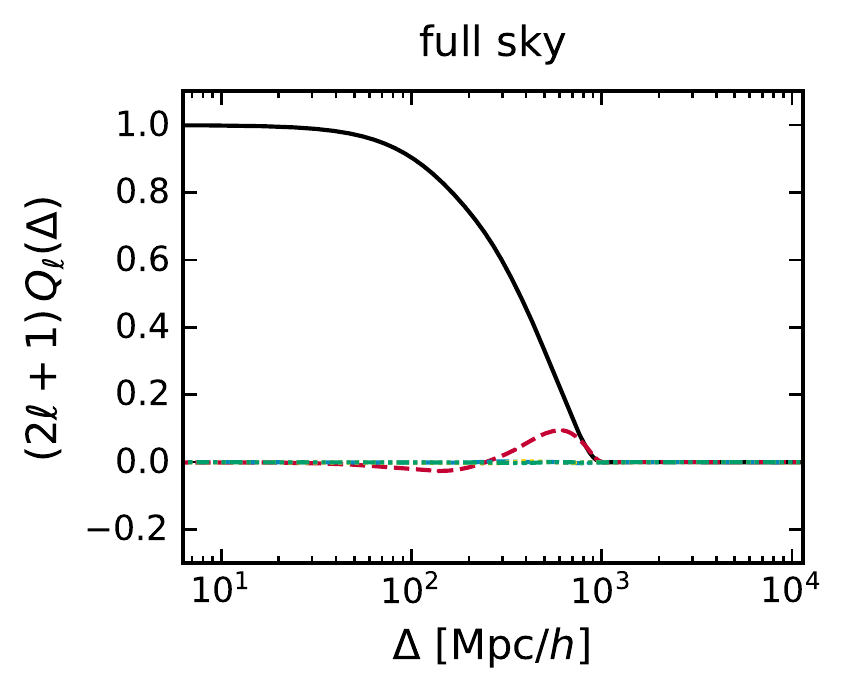}
    \hfill
    \includegraphics[width=.49\linewidth]{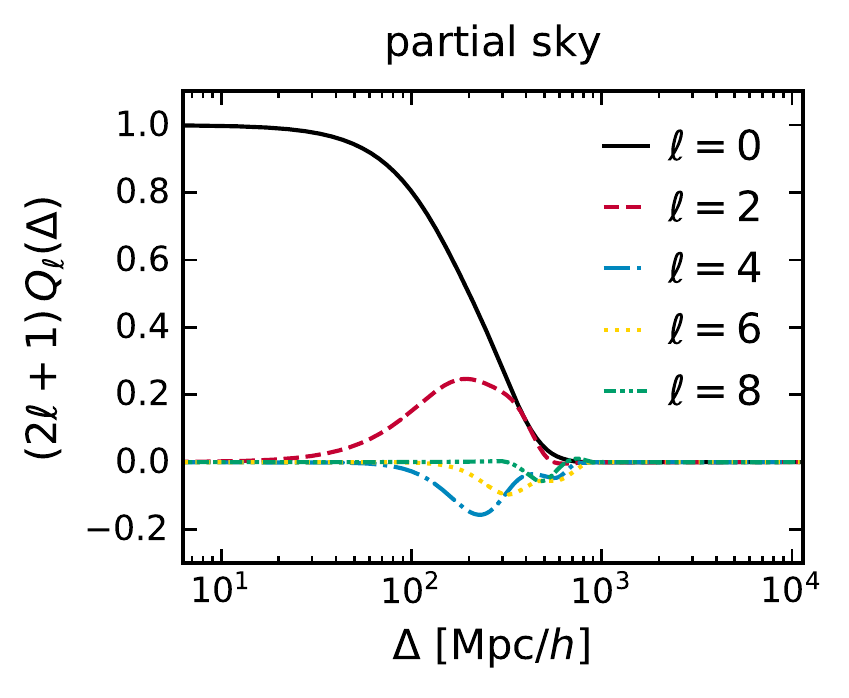}
    \caption{Survey window auto-correlation multipoles~$Q_\ell(\varDelta)$ normal{\is}ed to $Q_0(0) = 1$ in the full-sky (\textit{left column}) and partial-sky (\textit{right column}) set-ups. Each $\ell$-multipole is weighted by $(2\ell + 1)$ for visual~clarity.}
    \label{fig:window multipoles}
\end{figure}

Next, we determine the survey window auto-correlation multipoles~$Q_\ell(\varDelta)$ from FFTs of the power spectrum of a synthetic random catalogue. The number density field of the catalogue is first interpolated using the triangular-shaped cloud (TSC) scheme on a cubic grid with side length~$L = \SI{70}{\per\h\mega\parsec}$ and mesh number~$N_\textrm{mesh} = 768$. The large dimensions of the grid and the high mesh number allow us to compute the power spectrum across a wide range of scales without significant sample variance on very large scales or aliasing effects on very small scales. We then Hankel transform the power spectrum multipoles to $Q_\ell(\varDelta)$ using eq.~\eqref{eq:convolved power spectrum multipoles}. In figure~\ref{fig:window multipoles}, we show the multipoles~$Q_\ell(\varDelta)$ in both full-sky and partial-sky set-ups; in practice, only $Q_0(\varDelta)$ is needed for our Cartesian power spectrum analysis as we only consider the monopole for real-space clustering.

\begin{figure}
    \centering
    \includegraphics[width=.49\linewidth]{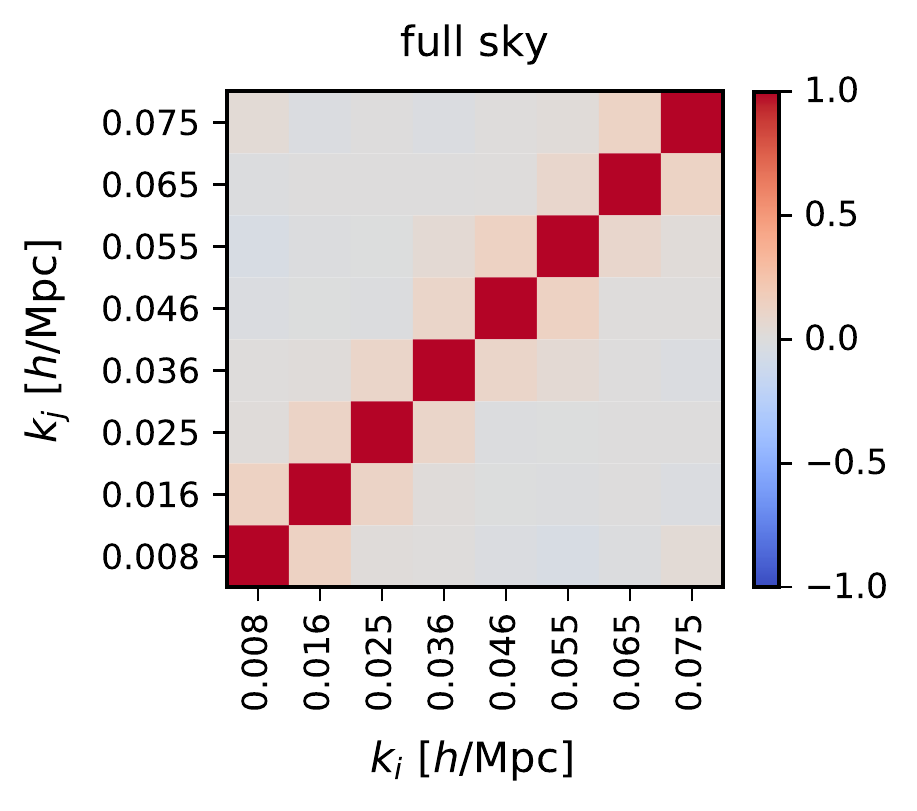}
    \hfill
    \includegraphics[width=.49\linewidth]{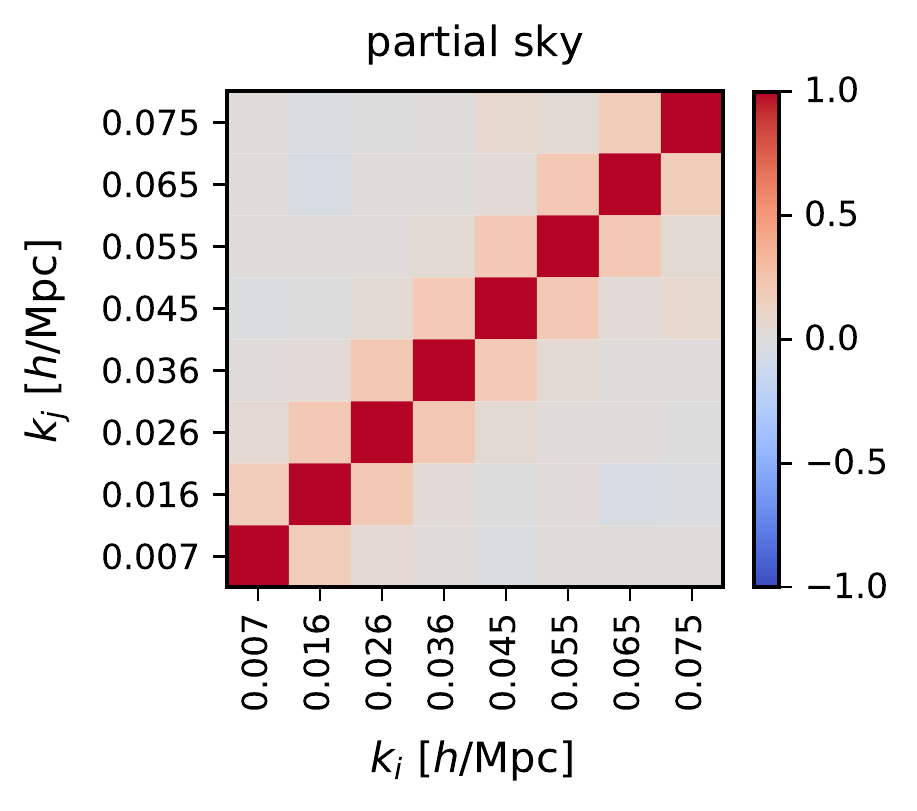}
    \caption{Estimated correlation matrices of the power spectrum monopole~$\est{P}_0(k)$ in $k$-bins up to $k_\max = \SI{0.08}{\h\per\mega\parsec}$ in the full-sky (\textit{left column}) and partial-sky (\textit{right column}) set-ups. The wave number representing each $k$-bin is the average over all Cartesian clustering modes in that bin.}
    \label{fig:covariance estimates}
\end{figure}

Lastly, we use $N_\textrm{rand} = 2500$ synthetic random catalogues of $50$~times the number density of the halo mock catalogues (i.e.~$\alpha = 0.02$) to obtain a fiducial covariance matrix estimate $\est{\mat{\Sigma}}_\fid$ for the binned power spectrum monopoles~$\est{P}_0(k)$. The use of unclustered random catalogues is justified as we restrict our analysis to linear scales, where correlation between measured Cartesian clustering modes is solely induced by the survey geometry rather than gravitational non-linearities. In figure~\ref{fig:covariance estimates}, we show the corresponding correlation matrices in the full-sky and partial-sky set-ups, where the $k$-bins are represented by the average mode wave number in each bin.

\subsection{Comparison of hybrid-basis and Cartesian power spectrum likelihoods}
\label{subsec:analysis results}

\begin{figure}
    \centering
    \includegraphics[width=.49\linewidth]{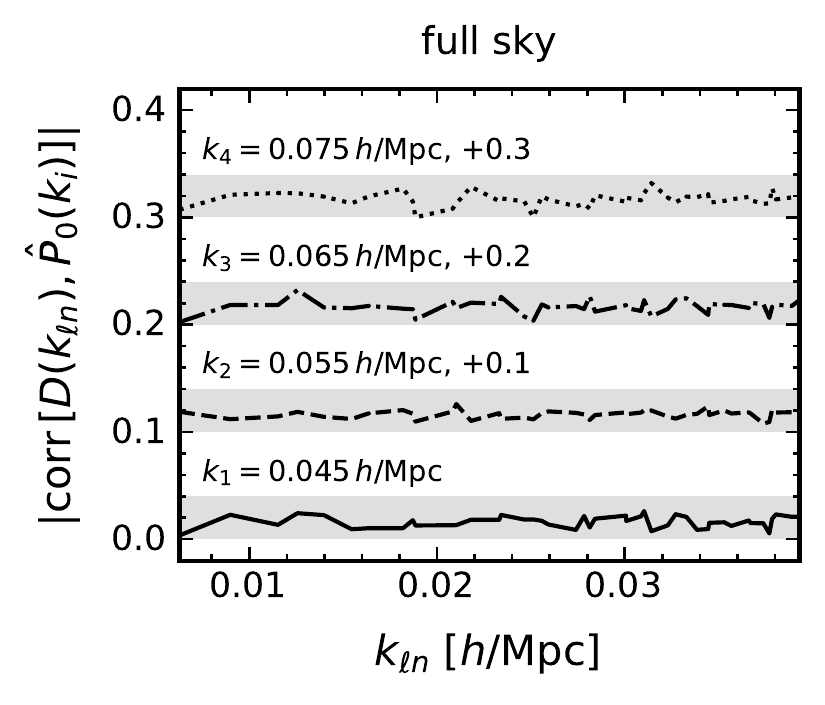}\hfill\includegraphics[width=.49\linewidth]{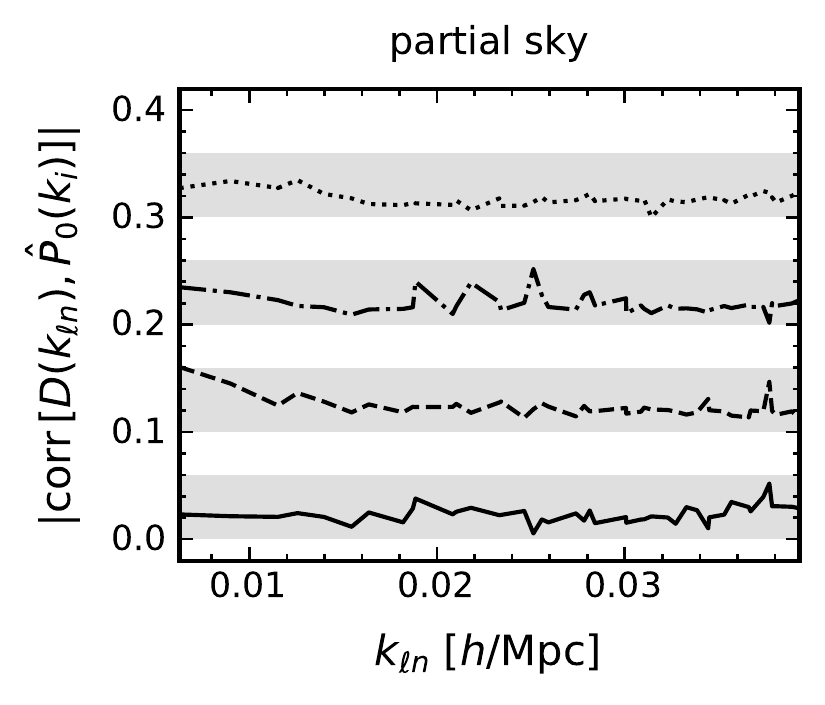}
    \caption{Absolute-value estimates of the cross-correlation~$\operatorname{corr}\big(\vb{D}, \est{\vb{P}}\big)$ between the spherical-basis data~$\vb{D}$ at wave numbers~$k_{\ell n} \leqslant k_\hyb$ and the Cartesian-basis data~$\est{\vb{P}}$ (power spectrum monopole only) in wave number bins~$k_i > k_\hyb$ in the full-sky (\textit{left column}) and partial-sky (\textit{right column}) set-ups. The shaded grey stripes are of width $0.04$ (\textit{left column}) or $0.06$ (\textit{right column}) in cross-correlation value. Plotted lines corresponding to the $i$-th wave number bin are shifted up by~$0.1 (i - 1)$ in cross-correlation value for visual clarity.}
    \label{fig:cross correlation}
\end{figure}

Before we perform parameter inference using the hybrid-basis likelihood, we check whether the correlation between low-$k$ spherical-basis data~$\vb{D}$ and high-$k$ Cartesian-basis data~$\est{\vb{P}}$ (power spectrum monopole only) is sufficiently weak so that eq.~\eqref{eq:hybrid-basis likelihood} is justified. To this end, we estimate the cross-correlation coefficients~$\operatorname{corr}\big(\vb{D}, \est{\vb{P}}\big)$ from the aforementioned $N_\textrm{rand} = 2500$ random catalogues in both the full-sky and partial-sky set-ups, as shown in figure~\ref{fig:cross correlation}. For SFB modes~$D_{\ell m n}$ of the same wave number $k_{\ell n}$ but different spherical orders~$m$, we average the absolute cross-correlation value over these equivalent modes. Indeed, the cross-correlation appears to be weak and there is no discernible evidence that particular SFB modes are more strongly correlated with the power spectrum monopole in any particular wave number bin. For the full-sky case, the cross-correlation coefficient is consistently below $0.04$ and for the partial-sky case, below $0.06$. Therefore we will treat $\vb{D}$~and~$\est{\vb{P}}$ as effectively independent under the assumption that the joint distribution of $\big(\vb{D}, \est{\vb{P}}\big)$ is multivariate normal.

Another issue related to the hybrid-basis likelihood is the inversion of the spherical-basis covariance matrix~$\mat{C}$. In our analysis, the spherical-basis data~$\vb{D} \in \C^{456}$ consist of 456~SFB modes, so the dimensions of $\mat{C} \in \C^{456 \times 456}$ are fairly large. As discussed in section~\ref{sec:hybrid basis approach}, this could render matrix inversion numerically unstable, and the likelihood function may diverge along some particular direction in parameter space. For the full-sky set-up, the spherical 2-point function~\eqref{eq:spherical 2-point function} is diagonal and thus $\mat{C}$ is well-conditioned; however, this is not the case for the partial-sky set-up as SFB modes become coupled. To deal with this issue, we follow the data compression procedure proposed in appendix~\ref{app:data compression}: we evaluate $\mat{C}(\theta_\fid)$ at fiducial cosmological parameters and obtain the compression matrix~$\mat{R} = \qty(\trans{\vb{e}}_1, \dots, \trans{\vb{e}}_{80}) \in \C^{80 \times 456}$ from $80$ eigenvectors~$\set{\vb{e}_j}$ of $\mat{C}(\theta_\fid)$ with the largest eigenvalues; we then apply the transformation~\eqref{eq:spherical-basis data compression} before evaluating the likelihood function.

\begin{figure}
    \centering
    \includegraphics[width=0.85\linewidth]{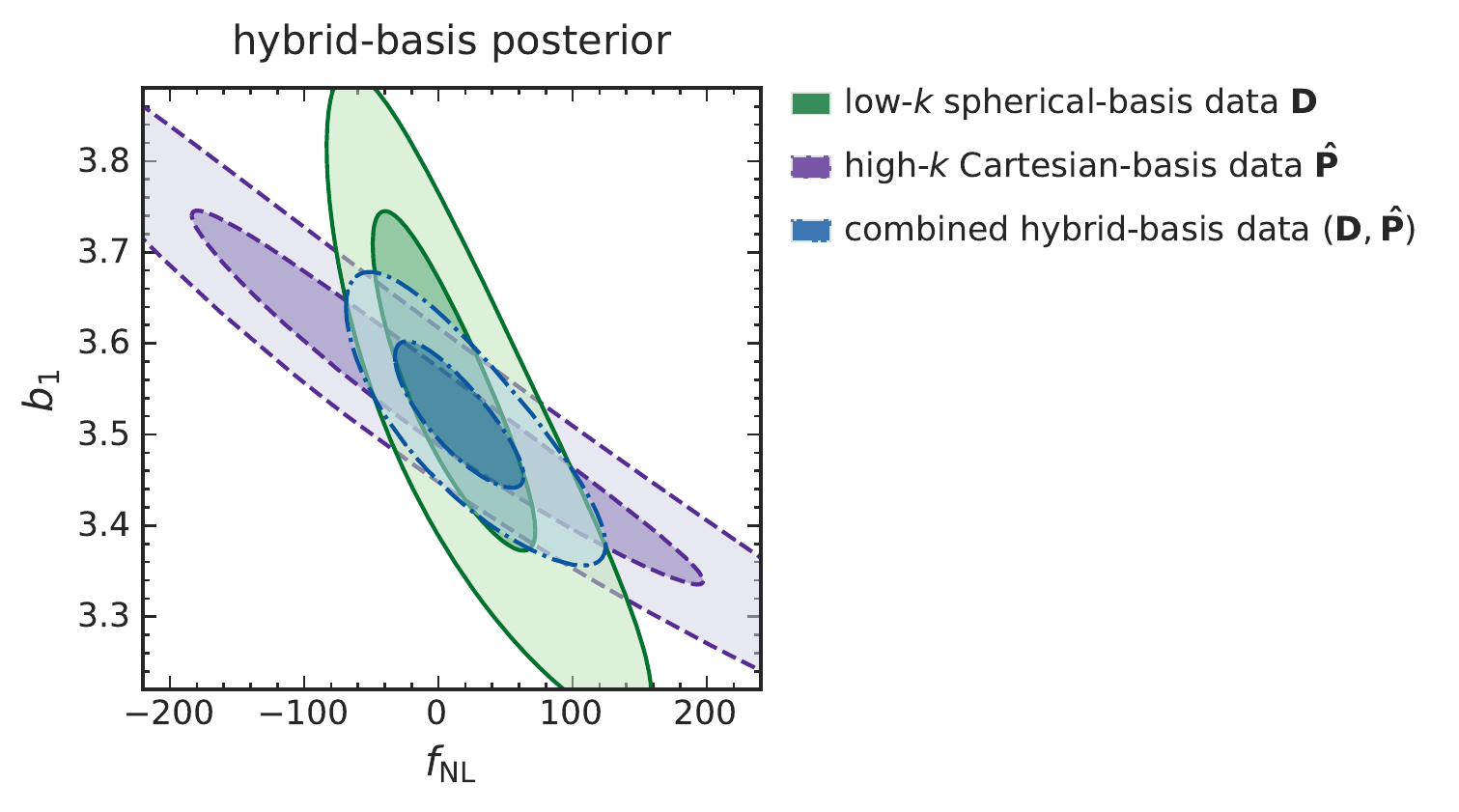}
    \caption{Joint posterior distribution of $\qty(\fNL, b_1)$ from the hybrid-basis likelihood~\eqref{eq:hybrid-basis likelihood} (dash-dotted blue contours) as well as its low-$k$ and high-$k$ components based on the spherical-basis data~$\vb{D}$ (solid green contours) and power spectrum data~$\est{\vb{P}}$ (dashed purple contours). In this example, results are marginal{\is}ed over from 24~mock catalogues with $\fNL = 0$. The inner and outer regions of each shaded contour set show the 1-$\sigma$ and 2-$\sigma$ credible bounds.}
    \label{fig:hybrid-basis likelihood components}
\end{figure}

Now that we have all the ingredients for computing the hybrid-basis and Cartesian-basis likelihoods, for each of our (non-)Gaussian halo mock catalogues in the full-sky or partial-sky set-up we infer local PNG~$\fNL$ and the scale-independent linear bias~$b_1$ jointly while keeping the background cosmology fixed. We choose uniform priors for both parameters so that the posterior distribution is simply proportional to the likelihood function. Throughout this section, measurements and inferred parameter constraints are presented as the marginal{\is}ed results over different sets of mock catalogues rather than the combined results which would have smaller uncertainties. This means that any presented data measurements have been averaged between equivalent mock catalogues and so are the logarithmic posterior distributions. As an example, we show in figure~\ref{fig:hybrid-basis likelihood components} the hybrid-basis posterior and its low-$k$ and high-$k$ components in the full-sky case. The results are marginal{\is}ed over 24~mock catalogues with $\fNL = 0$ by taking the average of the logarithmic posterior distributions.
The different orientations of the low-$k$ and high-$k$ posterior contours are mainly due to the different wave number ranges rather than differences in the spherical Fourier and Cartesian power spectrum analyses.

\begin{figure}
    \parbox[c][1.5\baselineskip]{\linewidth}{\centering\textbf{Full-sky parameter constraints}} \\
    \includegraphics[width=0.49\linewidth]{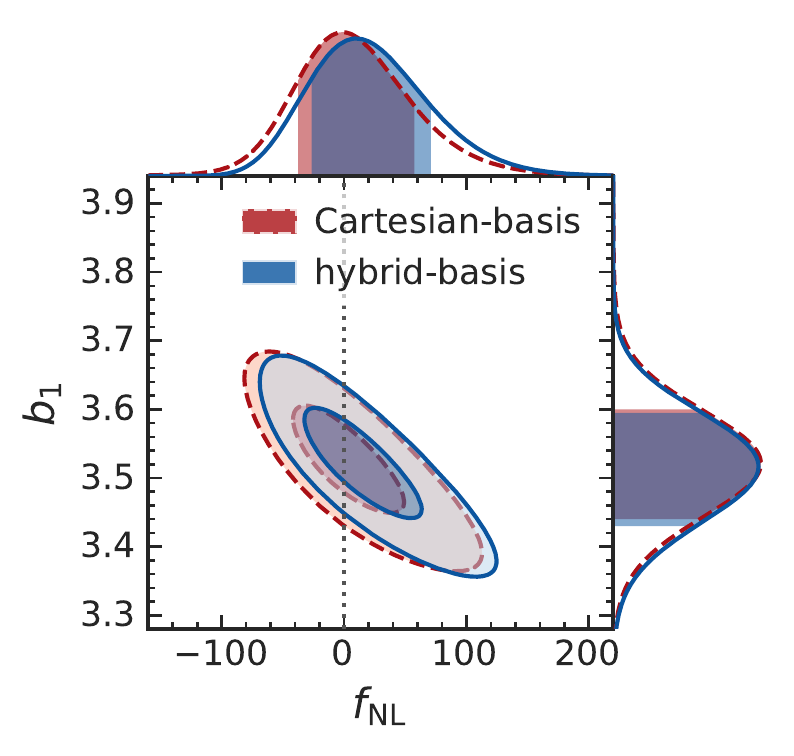}
    \hfill
    \includegraphics[width=0.49\linewidth]{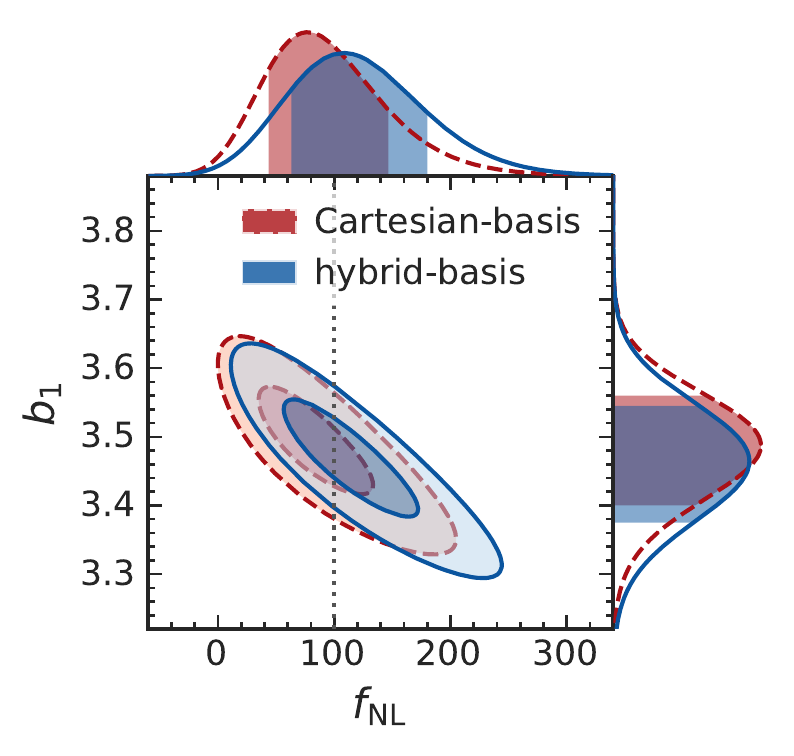}
    \caption{Posterior constraints on $\qty(\fNL, b_1)$ from the hybrid-basis and Cartesian power spectrum likelihood analyses of halo mock catalogues with $\fNL = 0$ (\textit{left column}) and $\fNL = 100$ (\textit{right column}) in the full-sky set-up. In the main panels, 1-$\sigma$ and 2-$\sigma$ credible regions of the joint posterior distribution are shown by the shaded contours, and the vertical dotted lines mark the true $\fNL$ values. The top and side panels show the marginal posterior distributions for $\fNL$ and $b_1$ respectively, with the shaded regions showing the 1-$\sigma$ credible interval. Hybrid-basis analysis results are coloured in blue and marked by solid lines, whereas Cartesian power spectrum analysis results are coloured in red and marked by dashed lines.}
    \label{fig:full-sky constraint}
\end{figure}

\begin{figure}
    \parbox[c][1.5\baselineskip]{\linewidth}{\centering\textbf{Partial-sky parameter constraints}} \\
    \includegraphics[width=0.49\linewidth]{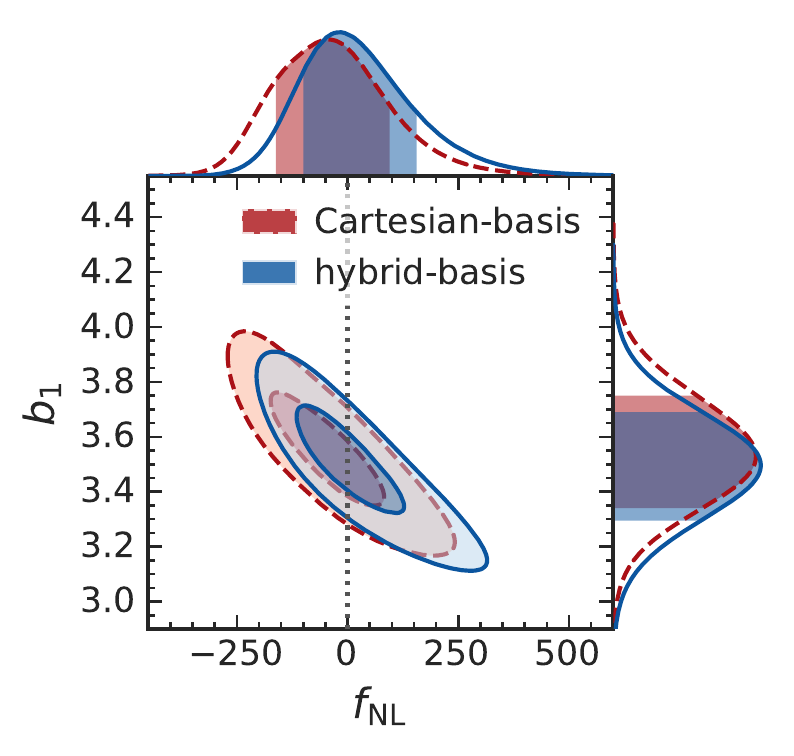}
    \hfill
    \includegraphics[width=0.49\linewidth]{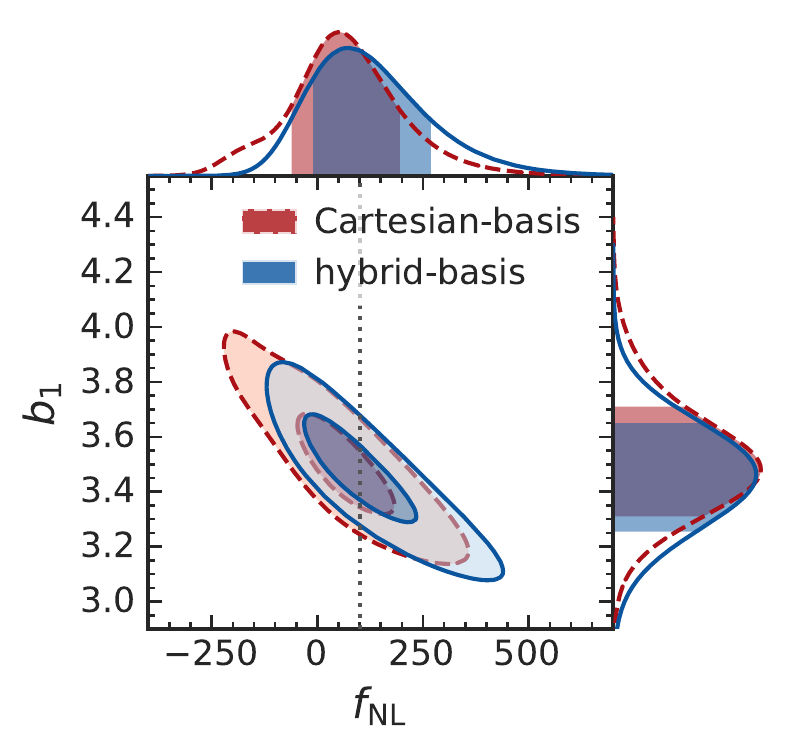}
    \caption{Posterior constraints on $\qty(\fNL, b_1)$ from the hybrid-basis and Cartesian power spectrum likelihood analyses of halo mock catalogues with $\fNL = 0$ (\textit{left column}) and $\fNL = 100$ (\textit{right column}) in the partial-sky set-up. In the main panels,  1-$\sigma$ and 2-$\sigma$ credible regions of the joint posterior distribution are shown by the shaded contours, and the vertical dotted lines mark the true $\fNL$ values. The top and side panels show the marginal posterior distributions for $\fNL$ and $b_1$ respectively, with the shaded regions showing the 1-$\sigma$ credible interval. Hybrid-basis analysis results are coloured in blue and marked by solid lines, whereas Cartesian power spectrum analysis results are coloured in red and marked by dashed lines.}
    \label{fig:partial-sky constraint}
\end{figure}

\begin{figure}
    \parbox[c][1.5\baselineskip]{\linewidth}{\centering\textbf{Full-sky comparison of inferred models}} \\[5pt]
    \includegraphics[width=0.49\linewidth]{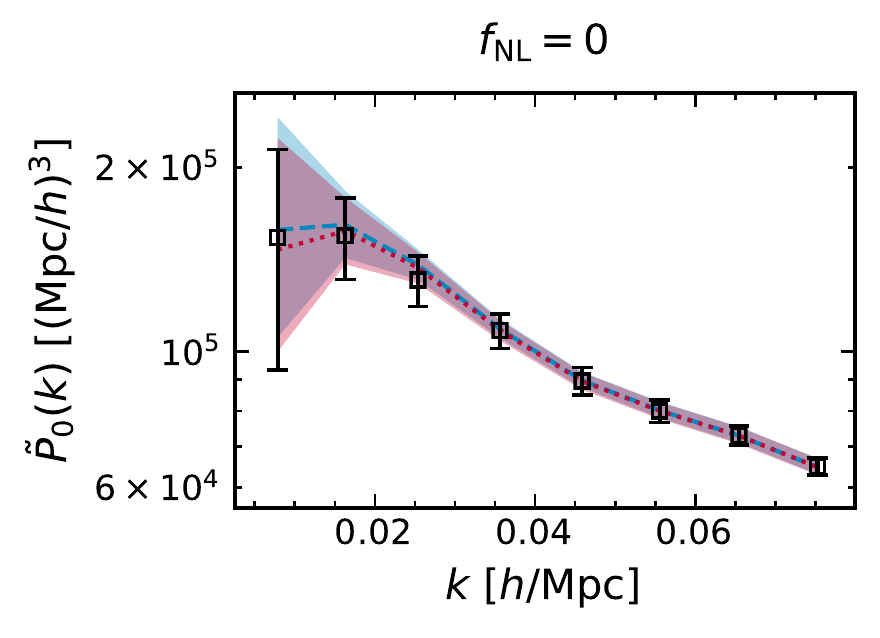}
    \hfill
    \includegraphics[width=0.49\linewidth]{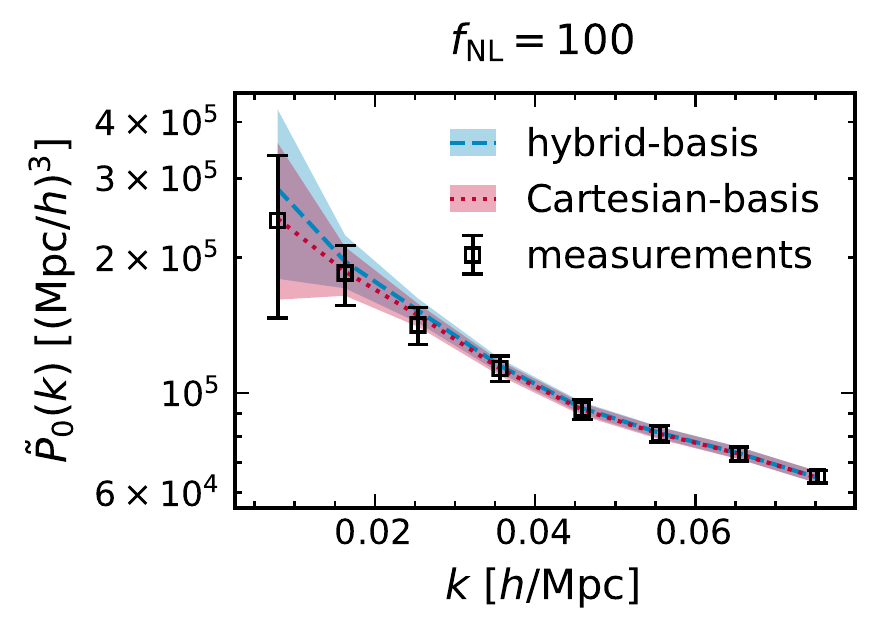}
    \caption{Comparison of the convolved power spectrum monopole model $\filter{P}_0(k)$ inferred from the joint posterior distribution of $\qty(\fNL, b_1)$ with the measurements averaged from halo mock catalogues with $\fNL = 0$ (\textit{left column}) and $\fNL = 100$ (\textit{right column}) in the full-sky set-up. Measurement uncertainties are obtained from the estimated covariance matrix and the shaded regions show the \SI{68}{\percent} credible interval of the inferred models from the hybrid-basis likelihood (dashed blue lines) and the Cartesian-basis power spectrum likelihood (dotted red lines).}
    \label{fig:full-sky comparison}
\end{figure}

\begin{figure}
    \parbox[c][1.5\baselineskip]{\linewidth}{\centering\textbf{Partial-sky comparison of inferred models}} \\[5pt]
    \includegraphics[width=0.49\linewidth]{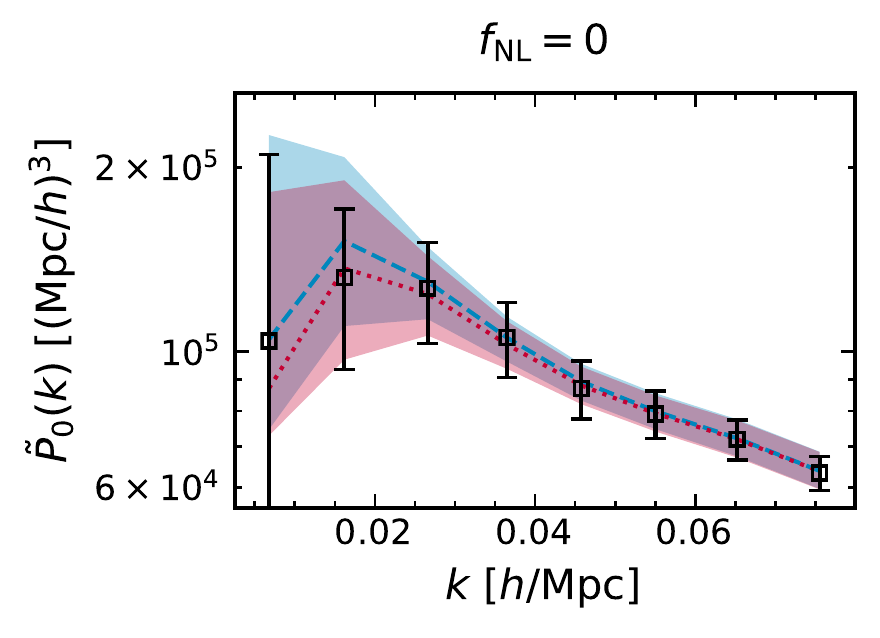}
    \hfill
    \includegraphics[width=0.49\linewidth]{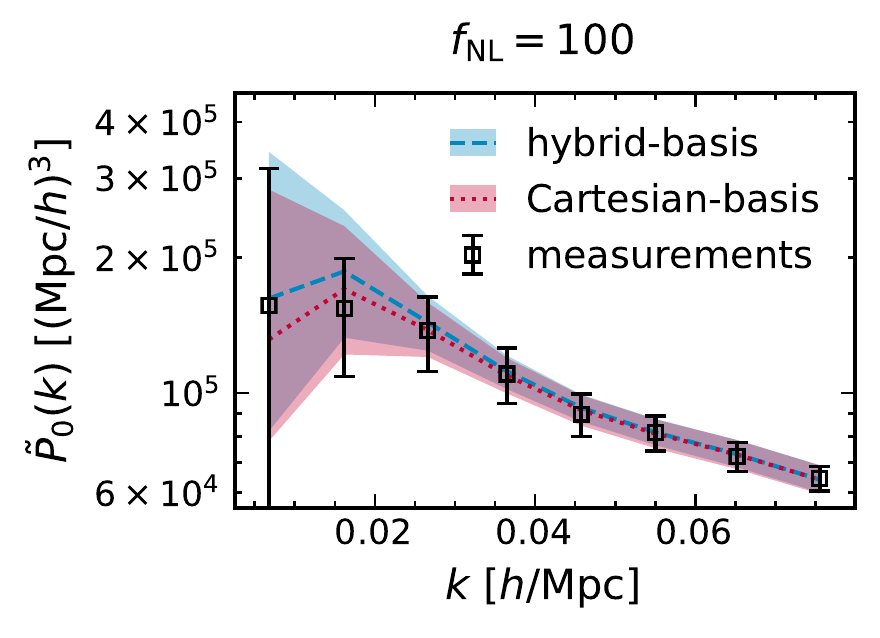}
    \caption{Comparison of the convolved power spectrum monopole model $\filter{P}_0(k)$ inferred from the joint posterior distribution of $\qty(\fNL, b_1)$ with the measurements averaged from halo mock catalogues with $\fNL = 0$ (\textit{left column}) and $\fNL = 100$ (\textit{right column}) in the partial-sky set-up. Measurement uncertainties are obtained from the estimated covariance matrix and the shaded regions show the \SI{68}{\percent} credible interval of the inferred models from the hybrid-basis likelihood (dashed blue lines) and the Cartesian-basis power spectrum likelihood (dotted red lines).}
    \label{fig:partial-sky comparison}
\end{figure}

In figure~\ref{fig:full-sky constraint}, we present the full-sky joint parameter constraints on $(\fNL, b_1)$ marginal{\is}ed over the 24~mock catalogues with $\fNL = 0$ and 20~mock catalogues with $\fNL = 100$. The results from the hybrid-basis and Cartesian power spectrum likelihoods are in good agreement.
Similarly, in figure~\ref{fig:partial-sky constraint}, constraints are shown for the same mock catalogues in the partial-sky set-up with the BOSS-like angular mask and the radial selection cut. As above, results from the hybrid-basis and Cartesian power spectrum analyses are statistically consistent.
To compare the best-fitting parameters from the different posterior distributions, we tabulate in table~\ref{tab:parameter constraints} the posterior median estimates for $\fNL$ and $b_1$ with uncertainties given by the \SI{68}{\percent} credible intervals of their marginal posterior distributions.
\begin{table}
    \centering
    \caption{Posterior median estimates of $\fNL$ and $b_1$ from the hybrid-basis and Cartesian power spectrum likelihood analyses in the full-sky and partial-sky set-ups, marginal{\is}ed over halo mock catalogues with $\fNL = 0$ and $\fNL = 100$. Uncertainties for both parameters correspond to the \SI{68}{\percent} credible interval of the marginal posterior distribution.}
    \vspace{\baselineskip}
    \bgroup
    \setlength{\tabcolsep}{0.8em}
    \renewcommand{\arraystretch}{1.1}
    \begin{tabular}{rrcc}
        \toprule[1.4pt]
        \multicolumn{2}{c}{\multirow{2}{*}{Mock catalogues}} & \multicolumn{2}{c}{Posterior median estimates of $\qty(\fNL, b_1)$} \\
        \cmidrule{3-4}
        & & Hybrid-basis analysis & Cartesian power spectrum analysis \\
        \midrule[1.4pt]
        \multirow{2}{*}{Full sky} & $\fNL = 0$ & $\qty(\measurement{18}{+53}{-44}\,, \measurement{3.51}{+0.08}{-0.08})$ & $\qty(\measurement{6}{+51}{-43}\,, \measurement{3.52}{+0.08}{-0.08})$ \\
        \cmidrule{2-4}
        & $\fNL = 100$ & $\qty(\measurement{117}{+63}{-54}\,, \measurement{3.46}{+0.09}{-0.09})$ & $\qty(\measurement{89}{+57}{-45}\,, \measurement{3.48}{+0.08}{-0.08})$ \\
        \midrule[0.7pt]
        \multirow{2}{*}{Partial sky} & $\fNL = 0$ & $\qty(\measurement{11}{+146}{-112}\,, \measurement{3.50}{+0.20}{-0.20})$ & $\qty(\measurement{-40}{+136}{-122}\,, \measurement{3.54}{+0.22}{-0.20})$ \\
        \cmidrule{2-4}
        & $\fNL = 100$ & $\qty(\measurement{111}{+158}{-120}\,, \measurement{3.46}{+0.20}{-0.20})$ & $\qty(\measurement{62}{+134}{-124}\,, \measurement{3.50}{+0.20}{-0.18})$ \\
        \bottomrule[1.4pt]
    \end{tabular}
    \egroup
    \label{tab:parameter constraints}
\end{table}
Finally, in figures~\ref{fig:full-sky comparison}~and~\ref{fig:partial-sky comparison}, we directly compare the window-convolved models of the power spectrum monopole inferred from the posterior distributions in the hybrid-basis and Cartesian power spectrum analyses with the measurements averaged over the different sets of mock catalogues. In both the full-sky and partial-sky set-ups, the recovered models from both analyses are in good agreement with the measurements, with the 1-$\sigma$ credible intervals of the inferred models comparable to the measurement uncertainties given by the estimated covariance matrix. However, it is worth noting that the measurement uncertainties are derived under the assumption that the power spectrum data follow the multivariate normal distribution, whereas the credible intervals of the inferred models are obtained from the non-Gaussian posterior distributions of $\fNL$~and~$b_1$.

\section{Discussion}
\label{sec:discussion}

With access to huge cosmic volumes, future galaxy redshift surveys have the potential to probe cosmological physics close to the horizon scale. On such large scales, upcoming missions such as DESI and \textit{Euclid} are forecast to constrain local PNG with uncertainties~$\sigma_{\fNL} \simeq 5$ competitive to the \textit{Planck} result, and relativistic effects in galaxy clustering can also be possibly detected~\cite{Font_Ribera_2014,Amendola_2018,Mueller_2018,Beutler_2020}. However, various large-scale survey systemics, if unaccounted for in the likelihood analysis, threaten to degrade or bias these parameter constraints.

In this work, we have reviewed redshift-space galaxy clustering that is commonly modelled by the anisotropic power spectrum~$P(k, \mu)$ or the equivalent Legendre multipoles~$P_\ell(k)$ derived in the distant-observer and global plane-parallel approximations with a fixed line of sight~$\los$. In practice, the line of sight~$\los$ varies across the survey volume, so the power spectrum multipoles are commonly estimated with the FFT-based Yamamoto estimator $\est{P}_\ell(k)$ derived in the local plane-parallel approximation. The discrepancy between the global plane-parallel prediction and the local plane-parallel estimator is known as the wide-angle effect, which makes a significant contribution to systematic errors on large scales when coupled with the survey window function. Although wide-angle corrections have recently been derived, at a more fundamental level, the standard power spectrum analysis is based on Fourier modes decomposed in the plane-wave basis, which forces the spherical geometry of survey observations to align with a Cartesian coordinate system.

A more natural description of redshift-space galaxy clustering on large scales is the spherical Fourier analysis based on the discrete SFB clustering modes~$D_\mu$, since many of the physical and observational effects affect clustering measurements parallel and transverse to the line of sight differently. In this work, we have extended previous works by refs.~\cite{Fisher_1995,Heavens_1995,Nicola_2014} to coherently include the AP effect, redshift evolution and scale-dependent galaxy bias on linear scales; further extensions to our model, such as the inclusion of relativistic corrections and scale-dependent linear growth rate in modified gravity theories, should be reasonably straightforward. Although the spherical Fourier analysis offers many advantages such as a clear separation between radial and angular components as well as being fully 3-dimensional (i.e.~no tomographic binning in redshift~$z$) as discussed in section~\ref{subsec:comparison and connections between analyses}, it is computationally expensive especially when confronted with huge data sets from future surveys and harder to relate to current models of non-linear galaxy clustering.

Inspired by the hybrid estimator approach used in CMB studies~\cite{Efstathiou_2004,Hinshaw_2007,Planck_2014,Planck_2016,Planck_2019}, we have proposed in this work an analogous hybrid-basis approach to analysing LSS observations: below some hybrid{\is}ation wave number~$k_\hyb$ chosen for a survey, anisotropic galaxy clustering can be accurately described by a spherical Fourier analysis using SFB modes; above~$k_\hyb$, we switch to the Cartesian power spectrum analysis. This approach has some major benefits: no geometric approximations are needed on large scales where a small number of clustering modes can be particularly affected by the survey geometry, and the likelihood directly constructed from SFB modes is exactly Gaussian with an analytically tractable covariance matrix; on smaller scales, the large number of clustering modes can be compressed into power spectrum multipoles, which are computationally fast to evaluate with FFTs and can be related to non-linear galaxy clustering models (e.g.~the TNS model for RSDs~\cite{Taruya_2010}), while the likelihood is now well approximated by a multivariate normal distribution thanks to the central~limit~theorem.

As a first step in demonstrating the applicability of the hybrid-basis approach, we have analysed real-space clustering statistics of halo mock catalogues from a series of $N$-body simulations with both Gaussian and non-Gaussian initial conditions. By performing likelihood analysis on the local PNG parameter~$\fNL$ and the scale-independent halo bias $b_1$, we have found that the hybrid-basis approach yields statistically very consistent results with those from the Cartesian power spectrum analysis. We expect that, when applied to more realistic scenarios that include RSD and light-cone effects, the hybrid-basis approach will outperform the standard analysis for multiple reasons: first, it is in the presence of anisotropic clustering around the line of sight that the plane-parallel approximations start to break down, unless one corrects for wide-angle effects; secondly, when the clustering measurements span a wide redshift range, the tomographic analysis of power spectrum in different redshift bins is less optimal than the fully 3-dimensional spherical Fourier analysis; thirdly, the spherical Fourier analysis is better suited for treating angular systematics in future wide surveys. In forthcoming works, we shall extensively test the hybrid-basis approach with data sets that include these additional effects; in these more complex scenarios, we will also consider the sensitivity of hybrid-basis analysis to the hybridisation scale~$k_\hyb$, and whether the low-$k$ and high-$k$ components can still be treated as effectively independent in the hybrid-basis likelihood and, if not, how they can be appropriately combined. In addition, ref.~\cite{Samushia_2019} has recently proposed a modified spherical Fourier--Bessel basis that is tailored for even more realistic survey geometries such as a spherical shell or cap, which could also be implemented in the hybrid-basis framework.

To serve further works in the future, we have released our public code \codename{harmonia} (online link in \cref{footnote:Harmonia}) as a Python package designed for both handling catalogue data and modelling clustering statistics in both spherical and Cartesian Fourier bases. Optimal weighting schemes for the spherical Fourier analysis~\cite[e.g.][]{Heavens_1995}, though not covered in this work, can be readily implemented using the code. In the future, we will also consider more sophisticated algorithms that can accelerate the SFB transform~\cite[e.g.][]{Leistedt_2012}.

\acknowledgments

MSW thanks Florian Beutler, Minas Karamanis and Pierros Ntelis for helpful discussions. MSW is supported by the University of Portsmouth Student Bursary. SA is supported by the MICUES project, funded by the European Union's Horizon~2020 research programme under the Marie Sk\l{}odowska-Curie Grant Agreement No.~713366 (InterTalentum UAM). DB acknowledges support from the European Union's Horizon 2020 research and innovation programme ERC (BePreSySe, grant agreement 725327) and from MDM-2014-0369 of ICCUB (Unidad de Excelencia Mar\'{i}a de Maeztu). RC is supported by the UK Science and Technology Facilities Council grant ST/S000550/1. Research at Perimeter Institute is supported in part by the Government of Canada through the Department of Innovation, Science and Economic Development Canada and by the Province of Ontario through the Ministry of Colleges~and~Universities.

This work makes extensive use of the Python package \codename{nbodykit} developed by Hand et al.~\cite{Hand_2018}. Numerical computations are performed on the Sciama High Performance Computing cluster which is supported by the Institute of Cosmology and Gravitation~(ICG), the South East Physics Network~(SEPnet) and the University of Portsmouth.

\appendix

\section{Computational complexity of the spherical Fourier analysis}
\label{app:computational complexity}

To determine how many discrete SFB modes up to some maximum wave number~$k_\max$ are present in a survey volume with boundary radius~$R$, we need to know the number of positive zeros~$u_{\ell n}$ of the spherical Bessel function~$j_\ell$ that satisfy $u_{\ell n} \leqslant k_\textrm{max} R$ (see eq.~\ref{eq:discrete radial wave numbers}). Using the asymptotic expansion~$u_{\ell 1} \sim \ell$ as $\ell \rightarrow \infty$, we estimate the maximum spherical degree to be
    \begin{equation}
        \ell_\textrm{max} \simeq k_\textrm{max} R \,.
    \end{equation}
By considering another asymptotic expansion~$u_{\ell n} \sim (n + \flatfrac{\ell}{2} - \flatfrac{1}{4}) \uppi$ for a fixed spherical degree~$\ell$ as $n \rightarrow \infty$~\cite{McMahon_1894}, we can estimate the maximum spherical depth by
    \begin{equation}
        n_{\textrm{max},\ell} \simeq \frac{k_\textrm{max} R}{\uppi} - \frac{\ell}{2} + \frac{1}{4} \leqslant \frac{k_\textrm{max} R}{\uppi} + \frac{1}{4} \,.
    \end{equation}
Since for each spherical degree~$\ell$ there are $(2\ell + 1)$ equivalent spherical orders~$m = -\ell, \dots, \ell$, the total number of SFB modes is bounded above by
    \begin{equation}
        N_\textrm{mode} = \sum_{\ell = 0}^{\ell_\textrm{max}} (2\ell + 1) n_{\textrm{max},\ell} \leqslant \qty(\frac{k_\textrm{max} R}{\uppi} + \frac{1}{4}) (k_\textrm{max} R + 1)^2 \sim \frac{(k_\textrm{max} R)^3}{\uppi} \,,
    \end{equation}
but this bound is in general far from being saturated.

For spherical Fourier transforms of the survey and synthetic catalogues, if we construct the SFB modes by direct summation (eq.~\ref{eq:SFB direct summation}), the number of computations is simply $(1+\alpha^{-1})N_\gal N_\textrm{mode}$ where each unit of computation is an evaluation of the spherical Bessel and harmonic functions. The calculation of the spherical couplings (eq.~\ref{eq:spherical couplings}) is more laborious: the number of angular coupling coefficients is
    \begin{equation}
        \qty[\sum_{\ell = 0}^{\ell_\textrm{max}} (2\ell + 1)]^2 \sim (k_\textrm{max}R)^4 \,,
    \end{equation}
but this can be reduced by employing symmetry relations between the spherical harmonics; on the other hand, for both radial and RSD couplings, the number of coupling coefficients is
    \begin{equation}
        \qty(\sum_{\ell = 0}^{\ell_\textrm{max}} n_{\textrm{max},\ell})^2 \leqslant \qty[(\ell_\textrm{max} + 1)\qty(\frac{k_\textrm{max} R}{\uppi} + \frac{1}{4})]^2 \sim \frac{(k_\textrm{max} R)^4}{\uppi^2} \,,
    \end{equation}
and the number of shot noise integrals (eq.~\ref{eq:shot noise 2-point function}) is similar. Finally, the infinite series~\eqref{eq:spherical 2-point function} requires \emph{at least} $N_\textrm{mode}$ terms for convergence, each of which is itself a product of the spherical coupling coefficients and the linear matter power spectrum.

Although the computational cost of angular coupling coefficients~$M_{\mu\nu}$ seems to be the highest, like shot noise they are independent of cosmology and thus need to be calculated only once for a given survey. In contrast, the radial and RSD couplings can change with the cosmological model if redshift evolution is to be taken into account, so their evaluations are likely to be the most expensive steps in a full likelihood analysis.

In this work, the maximum wave number for the spherical Fourier analysis is set to $k_\hyb = \SI{0.04}{\h\per\mega\parsec}$ and there are $456$~SFB modes in total with~$\ell_\max = 15$. On a single processor core, the computation time of each SFB mode~$D_\mu$ by direct summation is about $\SI{7e-5}{\second}$ per halo in our mock catalogues, or order of a day per mode for all galaxies in a DESI-like survey. The spherical coupling coefficients need to be computed once only for wave numbers up to $k_\textrm{trunc} = \SI{0.055}{\h\per\mega\parsec}$ (see discussions in sections \ref{sec:hybrid basis approach} and \ref{subsec:catalogue data}), since we fix the fiducial background cosmology; this takes about half an hour on 100~processor cores, and almost all the time is spent on the angular coupling coefficients~$M_{\mu\nu}$ (eq.~\ref{eq:angular couplings}) evaluated using the \codename{HEALPix} pixelation scheme with $N_\textrm{side} = 256$. A single evaluation of the spherical-basis likelihood~\eqref{eq:spherical-basis likelihood} takes just under a minute for the partial-sky set-up in section~\ref{sec:application}.\footnote{For the full-sky set-up where a number of simplifications can be made (see section~\ref{subsec:comparison and connections between analyses}), the covariance matrix $\mat{C}$ is diagonal, so the spherical-basis likelihood~\eqref{eq:spherical-basis likelihood} can be evaluated much faster.} If we were to extend to $k_\hyb = \SI{0.05}{\h\per\mega\parsec}$, there would be $978$~SFB modes with~$\ell_\max = 20$; the total computation time of both SFB modes and angular coupling coefficients would roughly double. Note that these figures are for reference only: more sophisticated numerical algorithms exist and can be incorporated in future work, as mentioned in section~\ref{sec:discussion}, and implementation in different programming languages can also result in different computation times.

\section{Data compression in the spherical Fourier analysis}
\label{app:data compression}

Because of the high computational cost associated with the spherical Fourier analysis, previous applications to galaxy survey catalogues~\cite[e.g.][]{Taylor_2001,Percival_2004} had to adopt data compression techniques, such as the Karhunen--Lo\`{e}ve transform which constructs optimal linear combinations of spherical clustering modes~\cite{Tegmark_1997b}, to make the analysis feasible. Indeed, the need for data compression is not only out of consideration for computational costs, but also for numerical stability. In this appendix, we consider a simple data compression method that deals with the latter problem.

In section~\ref{sec:hybrid basis approach}, we have noted the issue that when dimensions of a matrix~$\mat{C} \in \C^{N_\data \times N_\data}$ are sufficiently large ($N_\data \gg 1$), matrix inversion may not be numerically stable --- since one cannot determine the elements of $\mat{C}$ arbitrarily precisely, small perturbations to $\mat{C}$ can lead to spurious results for $\mat{C}^{-1}$. If $\mat{C}$ is a covariance matrix which must be positive-definite, the inversion procedure can introduce negative eigenvalues in $\mat{C}^{-1}$ unless it happens to be diagonal. This can affect both likelihood calculations as well as Fisher forecasts for cosmological parameters, as noted by ref.~\cite{Taylor_2001,Percival_2004,Yahia-Cherif_2020}. At the root of this phenomenon is the \emph{condition number}~$\varrho(\mat{C})$, which gauges the sensitivity of precision in numerically inverted $\mat{C}^{-1}$ to perturbations in the elements of $\mat{C}$; for a covariance matrix, it is given by~\cite{Golub_1996}
    \begin{equation}
        \varrho(\mat{C}) = \frac{\lambda_1}{\lambda_{N_\data}} \,,
        \label{eq:condition number}
    \end{equation}
where $\lambda_1 > \cdots > \lambda_{N_\data} > 0$ are its eigenvalues arranged in descending order.

For the covariance matrix~$\mat{C}$ of the SFB modes, its eigenvalues~$\lambda_i$ are typically of similar orders of magnitudes when $\mat{C}$ is close to being diagonal, as they are related to the power spectrum (eq.~\ref{eq:2-point function diagonal reduction}). However, when the SFB modes become correlated because of the spherical couplings coefficients, $\mat{C}$ is not diagonal and some eigenvalues are repelled towards zero which can become unstable upon matrix inversion. This hints at a solution based on the \emph{principal component analysis}~(PCA), where combinations of SFB modes corresponding to smaller eigenvalues are considered to have less cosmological information and thus discarded.

A practical method for data compression is proposed as follows. We first consider the covariance matrix~$\mat{C}(\theta_\fid)$ evaluated at fiducial cosmological parameters, with its eigenvalue--eigenvector pairs~$(\lambda_j, \vb{e}_j)$ arranged in descending order by eigenvalue. Next, we set an acceptable condition number~$\varrho$, and find the smallest eigenvalue~$\lambda_j = \lambda_J$ such that $\lambda_1 / \lambda_j \leqslant \varrho$. We can then define the fixed compression matrix
    \begin{equation}
        \mat{R} = \qty(\trans{\vb{e}}_1, \dots, \trans{\vb{e}}_J) \in \C^{J \times N_\data} \,,
    \end{equation}
which satisfies the orthonormality condition~$\mat{R} \trans{\mat{R}} = \mat{I}$. Finally, we replace $\vb{D} \mapsto \mat{R} \vb{D}$ and $\mat{C}(\theta) \mapsto \mat{R} \mat{C}(\theta) \trans{\mat{R}}$ in the spherical-basis likelihood~\eqref{eq:spherical-basis likelihood}.

\begin{figure}
    \centering
    \includegraphics[width=0.775\linewidth]{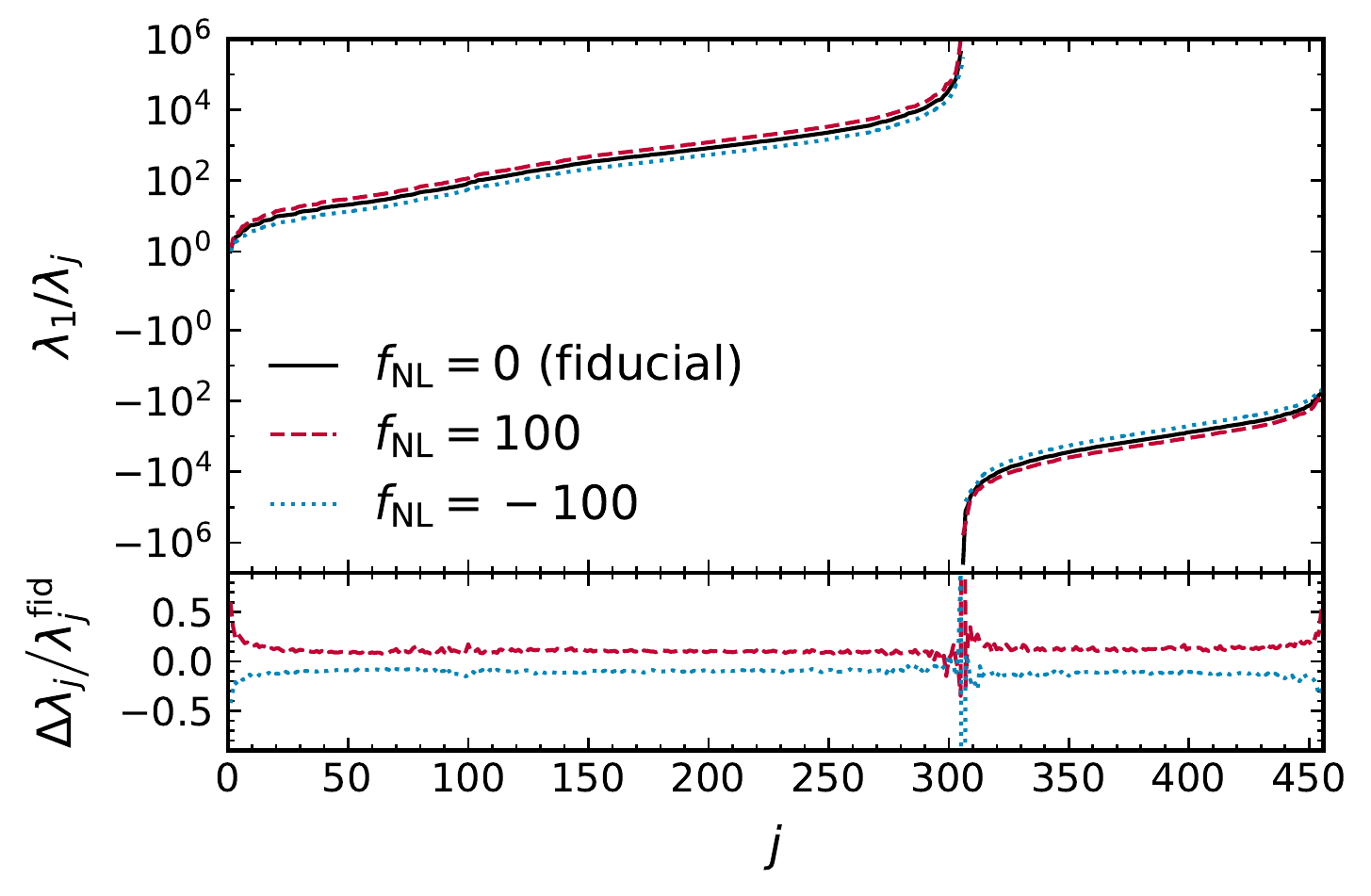}
    \caption{Variable condition number $\lambda_j/\lambda_1$ of the spherical-basis covariance matrix as a function of the index $j$ for eigenvalues $\lambda_j$ arranged in descending order. In the top panel, $\lambda_j/\lambda_1$ is shown for eigenvalues of the fiducial covariance matrix $\mat{C}_\fid$ with $\fNL = 0$ (solid black line) as well as covariance matrices $\mat{C}$ with $\fNL = \pm 100$ (dashed red line and dotted blue line respectively). In the bottom panel, the relative shift $\Delta \lambda_j$ in each eigenvalue compared to $\lambda_j^\fid$ of the fiducial covariance matrix $\mat{C}_\fid$ is shown. Note the sign change in eigenvalues around $j = 305$.}
    \label{fig:condition number}
\end{figure}

As an example, we consider the hybrid-basis likelihood analysis with $N_\data = 456$~SFB modes for the partial-sky set-up in section~\ref{subsec:analysis results}. The sky fraction is $\fsky \approx 0.2$ and together with the radial selection cut, the catalogue volume is only about \SI{18}{\percent} of the full-sky comoving sphere. Heuristically, the effective number of SFB modes is roughly $\SI{18}{\percent} \times N_\data \approx 82$, so we have kept only $80$~modes after data compression, corresponding to a conservative condition number $\varrho \approx 50$. In figure~\ref{fig:condition number}, we show the quantity~$\lambda_1/\lambda_j$ for all $N_\data = 456$ eigenvalues~$\lambda_j$ of the fiducial spherical-basis covariance matrix~$\mat{C}_\fid$ with $\fNL = 0$.
It is evident that there are many positive eigenvalues~$\lambda_j$ which are orders of magnitude smaller than the largest eigenvalue~$\lambda_1$, and almost a third of all eigenvalues are negative. We also check whether the eigenvalue composition alters significantly if a different covariance matrix $\mat{C}$, e.g.~with $\fNL = \pm 100$, is considered. The same figure shows that indeed the corresponding eigenvalues do not change very much except for the largest few and at the location where $\lambda_j$ switches sign; the number of negative eigenvalues remains almost the same. This suggests that our data compression method to stabil{\is}e the covariance matrix is robust.

\bibliographystyle{JHEP}
\bibliography{article}

\end{document}